\documentclass[11pt]{article}

\usepackage{arxiv}
\usepackage{hyperref}
\usepackage[utf8]{inputenc} 
\usepackage[T1]{fontenc}    
\usepackage{url}            
\usepackage{booktabs}       
\usepackage{amsthm,amsmath,amsfonts}
\usepackage{natbib}
\usepackage{float}
\usepackage{subcaption}
\usepackage{xcolor}
\usepackage{multirow}
\usepackage{nicefrac}       
\usepackage{microtype}      
\usepackage{graphicx}
\usepackage{anyfontsize}
\usepackage{afterpage}
\usepackage{rotating}
\usepackage{tabularx}
\usepackage{listings}
\usepackage{lstbayes}

\newcommand{\bolds}[1]{\boldsymbol{#1}}

\newcommand{\bD}{\bolds{D}}

\newcommand{\bI}{\bolds{I}}

\newcommand{\bQ}{\bolds{Q}}

\newcommand{\bbR}{\mathbb{R}}

\newcommand{\bW}{\bolds{W}}
\newcommand{\bx}{\bolds{x}}

\newcommand{\bY}{\bolds{Y}}

\newcommand{\Norm}{\mathcal{N}}
\newcommand{\bzero}{\mathbf{0}}

\newcommand{\bbeta}{\bolds{\beta}}

\newcommand{\bgamma}{\bolds{\gamma}}

\newcommand{\bphi}{\bolds{\phi}}

\newcommand{\lrnd}{\left(}
\newcommand{\rrnd}{\right)}
\newcommand{\lsq}{\left[}
\newcommand{\rsq}{\right]}
\newcommand{\lcur}{\left\lbrace}
\newcommand{\rcur}{\right\rbrace}
\renewcommand{\tilde}{\widetilde}

\definecolor{codegreen}{rgb}{0,0.6,0}
\definecolor{codegray}{rgb}{0.5,0.5,0.5}
\definecolor{codepurple}{rgb}{0.58,0,0.82}
\definecolor{backcolour}{rgb}{0.95,0.95,0.92}
\lstdefinestyle{mystyle}{
  backgroundcolor=\color{backcolour},   commentstyle=\color{codegreen},
  keywordstyle=\color{blue},
  numberstyle=\tiny\color{codegray},
  stringstyle=\color{codegreen},
  basicstyle=\ttfamily\scriptsize,
  breakatwhitespace=false,         
  breaklines=true,                 
  captionpos=b,                    
  keepspaces=false,                 
  numbers=left,                    
  numbersep=5pt,                  
  showspaces=false,                
  showstringspaces=false,
  showtabs=false,                  
  tabsize=2
}

\title{Spatial modelling of COVID-19 incident cases using Richards' curve: an application to the Italian regions}

\author{
Marco Mingione\\
    \scriptsize{Dpt. of Statistical Sciences}\\
    \scriptsize{University of Rome "La Sapienza"}\\
    \scriptsize{\texttt{marco.mingione@uniroma1.it}}
\And
    Pierfrancesco Alaimo Di Loro\\
    \scriptsize{Dpt. of Statistical Sciences}\\
    \scriptsize{University of Rome "La Sapienza"}\\
    \scriptsize{\texttt{pierfrancesco.alaimodiloro@uniroma1.it}}
\And
    Alessio Farcomeni\\
    \scriptsize{Dpt. of Economics and Finance}\\
    \scriptsize{University of Rome "Tor Vergata"}\\
    \scriptsize{\texttt{alessio.farcomeni@uniroma2.it}}
\And
Fabio Divino\\
    \scriptsize{Dpt. of Bio-Sciences}\\
    \scriptsize{University of Molise}\\
    \scriptsize{\texttt{fabio.divino@unimol.it}}
\And
    Gianfranco Lovison\\
    \scriptsize{Dpt. of Economics, Management and Statistics}\\
    \scriptsize{University of Palermo}\\
    \scriptsize{\texttt{gianfranco.lovison@unipa.it}}
\And
    Giovanna Jona Lasinio \\
      \scriptsize{Dpt. of Statistical Sciences}\\
    \scriptsize{University of Rome "La Sapienza"}\\
    \scriptsize{\texttt{giovanna.jonalasinio@uniroma1.it}}
\And
    Antonello Maruotti\\
     \scriptsize{Dpt. GEPLI}\\
    \scriptsize{Libera Università Maria Ss Assunta (LUMSA)}\\
    \scriptsize{\texttt{a.maruotti@lumsa.it}}
}

\begin{document}
\maketitle
\begin{abstract}
{ We introduce an extended generalised logistic growth model for discrete outcomes, in which a network structure can be specified to deal with spatial dependence and time dependence is dealt with using an Auto-Regressive approach.
  A major challenge concerns the specification of the network structure, crucial to consistently estimate the canonical parameters of the generalised logistic curve, e.g. peak time and height. Parameters are estimated under the Bayesian framework, using the {\texttt{ Stan}} probabilistic programming language. The proposed approach is motivated by the analysis of the first and second wave of COVID-19 in Italy, i.e. from February 2020 to July 2020 and from July 2020 to December 2020, respectively. We analyse data at the regional level and, interestingly enough, prove that substantial spatial and temporal dependence occurred in both waves, although strong restrictive measures were implemented during
  the first wave. Accurate predictions are obtained, improving those of the model where independence across regions is assumed.}

{\textbf{Keywords:} COVID-19, Conditional Auto-Regressive, \texttt{Stan}, generalised logistic growth.}
\end{abstract}

\section{Introduction}
\label{sec:intro}

Several approaches for modelling and forecasting COVID-19 incidence, prevalence, and related outcomes have been recently proposed. 
When high-quality data are available, compartmental models such as SIR/SEIR \citep{diak:13} are well known to describe the contagion dynamics satisfactorily and lead to scenario evaluation.
However, high-quality data are unlikely to be collected for the current epidemic, leading to the failure of most forecasts based on those approaches \citep{Ioannidis2020}.
Italy is not an exception. Indeed, public Italian data are gathered for mere descriptive and surveillance purposes, and present several issues that severely affect their quality.
Coherency, comparability, and consistency have been largely overlooked. Measurement errors and systematic biases are common. Data
collection systems are not standardised nor necessarily supported by proper digital or electronic devices and infrastructures.
Late notifications, corrections, and adjustments to the daily cases occur repeatedly.
In this context, direct data modelling seems to be a more viable option than compartmental models, allowing the researcher to deal with measurement error and restrict prediction to a short term \citep{greco:20, cabr:20, di2020nowcasting}.
An approach of this kind, as proposed by \cite{di2020nowcasting}, involves the specification of an appropriate parametric distribution for the available aggregated data (e.g., Poisson or Negative Binomial for counts), possibly meaningful predictors and offsets, and a logistic-type time trend. One of the limitations of this approach is that indicators in each area are modelled independently. That is clearly only a working assumption, as mobility have occurred across Italian regions, even during the hard lockdown of Spring 2020. Even sick people with COVID-19 have been sometimes transferred from one region to another. Furthermore, it is likely that regions close to each other culturally, economically, or geographically (e.g., sharing borders) present similar features as people experience similar climates, pollution and have similar lifestyles. 
For these reasons, this work aims to overcome this limitation by explicitly taking into account the spatial dependence \textit{across} regions and the temporal dependence \textit{within} regions.
We make this extension for different specifications of the generalized growth model of \cite{di2020nowcasting}, in a Bayesian framework. The Bayesian formulation by itself is already a notably additional advancement, regardless of model specification.
We report here that posterior summaries, in our experience, seem to be more stable compared to the maximum likelihood estimates, possibly due to difficulties in finding a global optimum for the likelihood of an inherently non-linear model. 
Furthermore, exploiting hierarchical models' flexibility in a Bayesian context, we replace the Negative Binomial assumption in \cite{di2020nowcasting} with the Poisson distribution. The still present over-dispersion and unobserved heterogeneity are accounted for by including observation specific random effects. 
If the random effect were assumed to be gamma-distributed, the corresponding marginal would indeed be a Negative Binomial and the method would be analogous to the original modelling framework. 
However, we rather consider normally distributed random effects on the log scale. Gaussianity allows for a more straightforward specification of prior information and inclusion of possible dependence structures in the process governing such effects. While temporal correlation is dealt with an \textit{Auto-Regressive} (AR) structure, spatial dependence is included by specifying a suitable \textit{Conditional Auto-Regressive} (CAR) prior, where the covariance matrix is identified using two possible networks: one based on geographic proximity and one built on historical data of transport exchanges between regions (taken from \citealp{della2020network}).
The advantage of introducing this dependence structure is twofold.
On the one hand, the resulting simultaneous model provably gives more accurate predictions than separate models for each region.
On the other hand, it can be expected that resulting parameter estimates for characteristics of interest (e.g., peak time and height) are also more accurate and less biased.
We separately evaluate the first and second wave of Sars-CoV-2 in Italy. 
According to \cite{bartolucci2021spatio}, we consider weekly incidence, even if observed cases are made available daily.
That is done in order to mitigate the issues with wild daily fluctuations due to late reporting. 
Even though we are aware that this does not solve the data issues, it sufficiently alleviates them, as testified by the smoother time series obtained at the weekly level. 
The remainder of the article is organised as follows: Sec.~\ref{sec:avadata} gives the necessary background, with some information about the available data
and the growth curve models proposed in a frequentist framework by \cite{di2020nowcasting}. The hierarchical structure of our model proposal is described in Sec.~\ref{sec:methodology}, while further details about the inclusion of the spatial and temporal dependence are given in Subsec.~\ref{subsec:meanspatial}. Sec.~\ref{sec:estimation} describes our Bayesian sampling strategy for approximating the posterior distribution of the model, and give insights on how to modify the CAR specification to gain in computational efficiency. Results of our modelling options for the first and second wave of COVID-19 in Italy are reported in Sec.~\ref{sec:results} and Subsec.~\ref{subsec:results}.
Concluding remarks can be found in Section \ref{sec:disc}. 

\section{Setup}
\label{sec:avadata}

Public data about COVID-19 in Italy are published every day by the Civil Protection Department,
since February 24th, 2020\footnote{{\texttt{GitHub}} repository: \href{https://github.com/pcm-dpc/COVID-19}{https://github.com/pcm-dpc/COVID-19}.}.
For each of nineteen regions and two provinces (Trento and Bolzano, forming the region of Trentino Alto Adige), these include 
(i) prevalence indicators (currently positive, Intensive Care Unit (ICU) occupancy, hospital occupancy) and
(ii) incidence indicators (e.g. newly diagnosed positives, deceased, new admissions to ICU, swabs, subjects tested).
For a more technical description refer to \cite{dicker2006principles, di2020nowcasting}.
For any of the incidence indicators in a given area, the number of new cases at time $t = 1, \dots, T$ can be obtained as the first difference of its cumulative counterpart as 
$
    Y_t = Y^c_t - Y^c_{t-1} 
$
where $Y_t$ and $Y_t^c$ are the number of new and cumulative cases at time $t$, respectively, and where we may assume $Y^c_0 = 0$ without loss of generality. 
Cumulative indicators present some peculiarities: they are monotone non-decreasing and their behaviour usually follows a logistic-type growth curve. 
Logistic curves were widely used to describe various biological processes \citep{werker1997modelling}.
Recently they have also been adapted in epidemiology and biostatistics for modelling the onset
and the spreading of epidemics \citep{hsieh2009richards, hsieh2009turning, hsieh2010pandemic}.

\cite{di2020nowcasting} proposed a modified Richards' curve \citep{richards1959flexible}, also known as the \textit{Generalised Logistic Function}, for modelling
cumulative incidence indicators. The generalised logistic function can accurately model various monotone processes
and include other widely-used logistic growth curves as special cases \citep{tsoularis2002analysis,gompertz1825xxiv}.
In \cite{di2020nowcasting} a parametric model is specified for region-specific incidence indicators (e.g., Poisson or Negative Binomial), where
the cumulative indicators are assumed to follow a five-parameters Richards' curve. 

The classical formulation of Richards' curve can be expressed as:
\begin{equation}\label{eq.StandRich}
    \begin{aligned}
        \Lambda_{\boldsymbol{\gamma}}(t) = b + \frac{r}{(1 + e^{h(p-t)})^s}, \qquad \boldsymbol{\gamma}^{\top} = [b, r, h, p, s]
    \end{aligned}
\end{equation}
where $b\in\mathbb{R}^+$ represents a lower asymptote (or baseline), $r\in \mathbb{R}^+$ is the distance between the upper and the lower asymptote, $h\in\mathbb{R}$ represents the growth rate, $p\in\mathbb{R}$ determines the peak position (when $s=1$ it corresponds to that), and $s\in\mathbb{R}$ is an asymmetry parameter regulating differences in the behaviour of the ascending and descending phase of the curve. 
Due to the monotone behaviour of cumulative incidence indicators, growth rate $h$ and asymmetry parameter $s$ are constrained to be positive.
However, a maximum of $b+r$ is foreseen for Richards' curve as $t\to\infty$. This implies that the virus would be eradicated at some point, an option which seems unlikely as for now. In order to allow the model to reach an endemic state in which there is a constant (hopefully, small) growth, we generalise Eq.~\eqref{eq.StandRich} by considering a linear trend on the baseline $b$:
\begin{equation}
\label{expRich}
    \Lambda_{\boldsymbol{\gamma}}(t) = b\cdot t + \frac{r}{(1 + e^{h(p-t)})^s}.
\end{equation}
This is similar to the use of an endemic parameter for the first differences as pursued in \cite{di2020nowcasting}. 

Assuming that the expected value of $Y_t^c$ (cumulative incidence indicator at time $t$) can be modeled with a (possibly modified) Richards' curve, i.e. $\mathbb{E}[Y^c_t] = \Lambda_{\boldsymbol{\gamma}}(t)$, 
the expected value for the innovation $Y_t$ can be straightforwardly obtained as:
\begin{equation}
\label{eqRespFun}
    \begin{aligned}
    \mathbb{E}[Y_t] &= \mathbb{E}[Y^c_t] - \mathbb{E}[Y^c_{t-1}] =  \Lambda_{\boldsymbol{\gamma}}(t) - \Lambda_{\boldsymbol{\gamma}}(t-1) = \lambda_{\bgamma}(t),
 \end{aligned}
\end{equation}
where:
\begin{equation}
\label{eq:diffRich}
\begin{aligned}
    \lambda_{\bgamma}(t) = 
     b + r\cdot\lsq \lrnd 1+ \exp\lcur h(p-t)\rcur\rrnd^{-s}-\lrnd 1+ \exp\lcur h(p-t+1)\rcur\rrnd^{-s}\rsq.
\end{aligned}
\end{equation}

\section{Generalised Logistic Growth Curve model with space-time dependence}
\label{sec:methodology}

Let $\mathbf{Y}_g=\lsq Y_{gt}\rsq_{t=1}^T$ denote the time-series of number of \textit{new} cases in area $g$, for $g = 1, \dots, G$, such that 
\begin{equation*}
            \bY=\lsq\mathbf{Y}_1^{\top},\mathbf{Y}_2^{\top},\dots,\mathbf{Y}_{G}^{\top}\rsq^{\top}.
\end{equation*}

The main assumption of our model is that $Y_{gt}$ arises from a  Poisson distribution with mean $\mu_{gt} = E_g \cdot m_{gt}$. This can be expressed as:
\begin{equation*}
    \begin{aligned}
            & Y_{gt}| \mu_{gt} \sim Pois(\mu_{gt})\\
        & \log(\mu_{gt}) = \log(E_g) + \log(m_{gt}), \quad g =1, \dots, G, \quad t = 1, \dots, T,
    \end{aligned}
\end{equation*}
where $\log(E_g)$ is an offset term that accounts for region-specific exposures levels.

When the offset is present, all other parameters impacting the overall rate become dimensionless, regardless of the scale of the corresponding region.
In other words, the term $m_{gt}$ can be interpreted as a relative measure of the risk of region $g$ at time $t$ with respect to the considered offset $E_g$. 

Different specifications of $m_{gt}$ lead to different models, each with its own characteristics. We decompose the log-risk in three main components:
\begin{equation}\label{eq:meantermspecgen}
    \log(m_{gt}) = \phi_{gt} +\log\lrnd\lambda_{\bgamma_g}(t)\rrnd + \boldsymbol{x}_{gt}^{\top} \bbeta,
\end{equation}
where $\phi_{gt}$ is a specific random effect for the $g$-th area at time $t$, $\lambda_{\boldsymbol{\bgamma}_g}(t)$ is a deterministic function denoting the general time trend, with possibly region specific parameters $\bgamma_g$ as for Eq.~\eqref{eq:diffRich}, and $\boldsymbol{x}_{gt}^{\top} \bbeta$ a linear predictor based on $K$ covariates with associated regression coefficients $\bbeta$.

\subsection{The Spatio-Temporal CAR model}
\label{subsec:meanspatial}
The observation-specific random effects $\lcur\phi_{gt}\,:\, g=1,\dots,G,\, t=1,\dots,T\rcur$ are included to account for unobserved heterogeneity in the data. At each time, possibly correlated random effects allow regional curves to deviate from their global average. Besides their presence corrects for the evident over-dispersion (with respect to the Poisson assumption) present in the time-series.

These random effects can either be completely independent, present temporal dependence, spatial dependence, or spatio-temporal dependence. 
In a Bayesian framework, the covariance structure can be induced hierarchically by specifying a suitable prior on the complete set of random effects. 
In order to simplify the formulation of the random effects prior, we collect all of them in a set of time-varying vectors $\bphi_t=\lsq\phi_{1t},\dots,\phi_{Gt},\rsq^\top,\; t=1,\dots,T$.  

Spatial dependence at each time point can be introduced by using a CAR prior \citep{besag1974spatial} over some network, that under Gaussianity produces a so-called \textit{Gaussian Markov Random Field} (GMRF, \cite{rue2005gaussian}). 
This approach falls into the wide range of methods related to \textit{disease mapping} (see \cite{waller2010handbook} and \cite{lawson2018bayesian} for a review). Such CAR prior specification allows incorporating the undeniable spatial correlation at the second level of the model hierarchy, avoiding analytical complications inherent in modelling spatial correlation within non-Gaussian distributions with inter-related mean and variance structures \citep{gelfand2010handbook}.
This form of dependence is valid on discrete domains arranged over a network, where neighbouring relationships are determined by an adjacency matrix $\bW$ (possibly weighted). The matrix $\bW = [w_{ij}]$ is a $G\times G$ symmetric matrix with all diagonal elements equal to $0$ (as no region/area/unit is its own neighbour), and where off-diagonal elements $w_{ij}$ are greater than $0$ if and only if areas $i$ and $j$ are connected ($i\sim j$): the larger the connection strength $w_{ij}$, the closer the two random effects are pulled together. 
The original expression of this prior starts from the consideration of the full conditional of each random effect given all the others. For the generic $t\in\lcur 1,\dots,T\rcur$, the full conditional $\phi_{gt}|\bphi_{-gt},\;g=1,\dots,G$ has mean equal to the weighted combination of the random effects in its neighbourhood:
\begin{equation*}
    \phi_{gt}|\bphi_{-gt}\sim\Norm\lrnd\sum_{j=1}^Gw_{gj}\phi_{jt}\,,\,\sigma^2\rrnd,\quad \forall\;g=1,\dots,G,
\end{equation*}
where $\bphi_{-gt}=\lsq\phi_{1t},\dots,\phi_{(g-1)t},\phi_{(g+1)t},\dots,\phi_{Gt},\rsq^\top$ and  $\sigma^2$ is the overall variance of the random effect. This induces smooth variations over close regions, as determined by $\bW$.
Following Brook's lemma, for a fully connected graph (i.e. with no ``islands''), this local specification implies a very specific global multivariate prior on the vector $\bphi_{t}$, centered at $\bzero$ and with precision matrix $\bQ$ which depends on the network structure. 
Under row-wise normalization of the weights in $\bW$, and introducing a spatial smoothing parameter $\alpha$, this global prior can be expressed as:
\begin{equation}
\label{eq:CarModel}
    \bphi_t \sim\Norm_{G}\lrnd\bzero,\,\sigma^2\cdot\bQ(\alpha, \bW)^{-1}\rrnd,\quad \forall\;t=1,\dots,T,
\end{equation}
where $\bQ(\alpha, \bW) = \lrnd\bD-\alpha\bW\rrnd$ and the matrix $\bD$ is a diagonal matrix containing the row sums of the weights of each region on the diagonal. This simply ensures that the weights of each region are properly normalized over all its neighbours (i.e. the row-wise normalization).
We denote by $\alpha$  the spatial smoothing parameter, which actually regulates the amount of spatial dependence: values close to $0$ approximate independence (no impact of $\bW$) and values close to $1$ strong spatial dependence (full impact of $\bW$).
This spatial CAR expression allows the introduction of spatial dependence among random effects belonging to the same time-point. Nevertheless, we cannot neglect the evident temporal correlation which characterizes such kind of data. 

Hence, following the original work by \cite{rushworth2014spatio}, we induce such dependence by imposing a temporal AR(1) structure over the vectors $\lcur\bphi_t\rcur_{t=1}^T$.
This yields a spatio-temporal CAR model, also known as CAR-AR, whose only difference in this work from the original version is that, instead of the mixed specification of \cite{leroux2000estimation}, we here consider the typical CAR of \cite{besag1974spatial}.
This amounts to the following prior for the collection of time-varying spatial vectors:
\begin{equation*}
\begin{aligned}
&\bphi_1 \sim\Norm_{G}\lrnd\bzero,\,\sigma^2\cdot\bQ(\alpha, \bW)^{-1}\rrnd\\
&\bphi_t \sim\Norm_{G}\lrnd\rho\cdot \bphi_{t-1},\,\sigma^2\cdot\bQ(\alpha, \bW)^{-1}\rrnd, \quad t=2,\dots,T\\
\end{aligned}
\end{equation*}
where $\rho\in\lrnd -1, 1\rrnd$ is the temporal auto-regressive parameter governing the amount and direction of temporal dependence. The hyperparameters governing the chosen priors can be ascribed typical hyperpriors as follows:
\begin{equation*}
    \begin{aligned}
    & \alpha \sim \mathcal{B}e(0.5, 0.5) 
    \qquad \qquad \rho \sim \mathcal{U}nif(-1, 1) 
    \qquad \qquad  \sigma^2\sim \mathcal{IG}\lrnd 2,\,2\rrnd,
    \end{aligned}
\end{equation*} 
where the latter has been coded as $\sigma^2=\frac{1}{\tau^2}$ with $\tau^2\sim \mathcal{G}a\lrnd 2,\,2\rrnd$ for favouring the estimation process.

\subsection{The logistic growth trend}
In Sec.~\ref{sec:avadata}, we argue that we want to model the general trend of COVID-19 counts (of positives) in a single outbreak by using the first differences of the Richards' curve as in \cite{di2020nowcasting}. 
The first differences in Eq.~\eqref{eq:diffRich} do not present an elegant expression and are slightly cumbersome to work with. Since data are collected at equally spaced time intervals, we propose to linearly approximate $\lambda_{\bgamma}(t)$ with the derivative of the Richards' curve, as follows:
\begin{equation}
\label{eq:diffRichLin}
\begin{aligned}
    \lambda_{\bgamma} (t) &\approx \tilde{\lambda}_{\bgamma} (t)=\frac{d}{d t}\Lambda_{\bgamma}(t)\cdot \Delta t = b + r\cdot s\cdot h\cdot\exp\lcur h\cdot (p-t)\rcur\cdot\lrnd 1+ \exp\lcur h(p-t)\rcur\rrnd^{-(s+1)},
\end{aligned}
\end{equation}
where $\Delta t =1$. In our implementation, we initially considered both the exact and the linearised version of Eq. \eqref{eq:diffRich} and \eqref{eq:diffRichLin}, respectively.  In the final results, differences were negligible, but the latter provided improved numerical stability and convergence of the chains. Thus, we decided to stick to this version, which is also used to produce the final results included in Subsec.~\ref{subsec:results}.

The expression in Eq.~\eqref{eq:meantermspecgen} implicitly considers a very highly parametrised model, where each region is allowed its on vector of parameters $\lcur \bgamma_g\rcur_{g=1}^G$, hence its own Richards' curve, to drive the trend of the regional outbreaks.  
In the sequel, we alternatively envision the existence of one common single Richards' curve governing the spread of the epidemic in all the Italian regions, which then deviate from this \textit{global average} as an effect of specific characteristics (observed or unobserved). This is obtained as a particular case of the former, where $\bgamma_g=\bgamma,\;\forall\,g\in\lcur 1,\dots,G\rcur$.

There is an essential difference between these two specifications, especially in terms of the role of the space-time random effects. The first one is a local model, where the random effects represent temporal variations of the mean underlying each regional counts-series from the region-specific trend. In the second case, they instead represent the spatio-temporal deviations of each region's means, at each time, from the common curve.
From a dependence interpretation standpoint, in the first case we are assuming that if region $g$ and region $j$ are connected, when region $g$ deviates from its trend $\lambda_{\bgamma_g}(\cdot)$, then region $j$ will likely have similar deviation from its own trend $\lambda_{\bgamma_j}(\cdot)$ as well.
In the second case, we are assuming that if region $g$ and region $j$ are connected, when region $g$ deviates from the general trend $\lambda_{\bgamma}(\cdot)$, then region $j$ will also deviate the general trend similarly.

Let us recall that the parameters $\lcur b, r, h, s\rcur$ governing the differences of the Richards' curve are constrained on the positive domain $\bbR^+$. In order to favour the elicitation of diffuse priors and the Bayesian estimation process, these have been parametrised on the log-scale as $\lcur\log(b), \log(r), \log(h), \log(s)\rcur$. The first two have been assigned a $\Norm(0, 100)$ prior, while the last two a $\Norm(0,1)$ one.

The parameter $p$, differently from the others, belongs to the whole real line $\bbR$ and, more importantly, is not dimensionless. Its magnitude is indeed related to the dimension of the analysed time window. It can be loosely interpreted as the lag-phase of the outbreak (for $s=1$ it represents precisely the point of maximum of the curve), which is the point in time when the exponent $h\cdot (p-t)$ becomes negative. It is not well-identified for varying $s$, and hence it has been given a $\Norm\lrnd T/2,\, T/(2\cdot 1.96)\rrnd$ prior to help it move inside the observed time interval (included in $[0, T]$ with $95\%$ probability).

\subsection{The linear predictor}

The linear predictor $\bx_{gt}^{\top}\bbeta$ describes the effect of covariates on the log-risks. Since the \textit{dimension} of the region is already accounted for in the offset term, such covariates shall account for exogenous factors that affect the spread of the virus, or the ability to detect the infected people in each area at different times. In practice, this term shall represent all the meaningful observed heterogeneity between regions and within regions over time.

For instance, the population density is a region-specific and constant over time feature that can likely impact on the rate of infection. This covariate has been considered in the recent work by \cite{jalilian2021hierarchical} and proved to be valid both for explanation and prediction purposes. Another interesting variable to study may be the number of daily swabs. Its inclusion accounts for the effort in detecting positive cases carried out by a region at a specific time. 

If $K$ covariates are considered, then the vector $\bx_{gt}^{\top}$ is associated to a $(K \times 1)$ vector of coefficients $\bbeta$. In our Bayesian machinery, this vector is assigned a multivariate Normal prior with independent components $\Norm_K\lrnd \bzero, 100\cdot \bI_K \rrnd$, which corresponds to a fairly diffuse prior considering the log-linear link.

It is here important to highlight that we are not including the intercept in the linear predictor. In the case of region-specific Richards' curves, the intercept is implicitly defined by the parameters $b_g$ and $r_g$ already, and its inclusion would introduce a non identifiable parameter and jeopardize proper convergence of the estimation algorithms.
In the case of a common single Richards' curve, one may want to include region-specific intercepts $\lcur\beta_{0g}\rcur_{g=1}^G$. These would have the effect of moving the whole region-specific curve up or down with respect to the \textit{global average}, again accounting for unobserved heterogeneity among regions. However, the goal is to have this heterogeneity  explained by the spatio-temporal random effects $\phi_{gt}$: the inclusion of such individual intercepts would add an unwanted player in the game and make the interpretation of the final results intricate.


\section{Estimation}\label{sec:estimation}

\begin{lstlisting}[language=Stan, label = sparsecar, caption = Core of the \texttt{Stan} code for the sparse implementation of the spatio-temporal CAR-AR., float=htpb]
functions {
    // Sparse computation of the log-posterior contribution of the single spatial vector phi_t (phi) given the previous value phi_{t-1} (phi_old)
    real sparse_carar_lpdf(vector phi, vector phiOld, real rho, real tau, real alpha, 
    int[,] W_sparse, vector W_weight, vector D_sparse, vector lambda, int n, int W_n) {
        row_vector[n] phit_D;
        row_vector[n] phit_W;
        vector[n] ldet_terms;
        vector[n] phiNew = phi-rho*phiOld;
        phit_D = (phiNew .* D_sparse);
        phit_W = rep_row_vector(0, n);
        for (i in 1:W_n) {
            phit_W[W_sparse[i, 1]] = phit_W[W_sparse[i, 1]] + W_weight[i]*phiNew[W_sparse[i, 2]];
            phit_W[W_sparse[i, 2]] = phit_W[W_sparse[i, 2]] + W_weight[i]*phiNew[W_sparse[i, 1]];
        }
    
      for (i in 1:n) ldet_terms[i] = log1m(alpha * lambda[i]);
      
      return 0.5*(n*log(tau) + sum(ldet_terms) - tau*(phit_D*phiNew - alpha*(phit_W*phiNew)));
  }
}

data {
  int<lower=0> nTimes;    // Number of times
  int<lower=0> nReg;      // Number of regions
  
  int W_n;                // Number of adjacent region pairs
  int W_sparse[W_n, 2];   // adjacency pairs
  vector[W_n] W_weight;   // Connection weights
  vector[Nreg] D_sparse;  // diagonal of D
  vector[Nreg] lambda;    // eigenvalues of invsqrtD * W * invsqrtD
}

parameters {  
    vector[Nreg] phi[Ntimes];
    real<lower=0, upper=1> alpha;
    real<lower=-1, upper=1> rho;
}

model {
    alpha ~ beta(0.5, 0.5);
    phi[1] ~ sparse_carar(zeros, rho, tau, alpha, 
                            W_sparse, W_weight, D_sparse, lambda, Nreg, W_n);
    for (i in 2:Ntimes){
            phi[i] ~ sparse_carar(phi[i-1], rho, tau, alpha, 
                            W_sparse, W_weight, D_sparse, lambda, Nreg, W_n);
        }
}
\end{lstlisting}

Estimation has been carried out using \texttt{Stan} probabilistic programming language\footnote{Webpage at \url{https://mc-stan.org/}}, which is a software for statistical modeling and high-performance statistical computation \citep{carpenter2017stan, Stan}. 
It interacts with \texttt{R} and can be called directly from \texttt{RStudio} \citep{allaire2012rstudio} through the \texttt{rstan} package \citep{Rstan}. 
Among its many capabilities, it allows to get full Bayesian inference by drawing from the posterior density by a specific Markov Chain Monte Carlo (MCMC) sampling method known as Hamiltonian Monte Carlo (HMC, \cite{betancourt2017conceptual}). The HMC techniques provide an efficient sampling scheme based on the simulation of Hamiltonian dynamics for approximating the target distribution \citep{neal2011mcmc}. Its functioning relies on the analogy between the parameter value and the trajectory of a fictitious particle subject to a potential energy field, preserving its total energy (the Hamiltonian), obtained as the sum of potential and kinetic energy. 
In practical terms, given the chain's current value and the corresponding log-density, it picks the new value of the chain by proposing a random shift in the log-density value and then moving arbitrarily far away from that point along the corresponding contour line.  
The latter allows for fast and complete exploration of the whole density that does not negatively affect the acceptance rate, minimizing the risk of wasting time (or even getting stuck) in local high-density areas.
Unlike the Metropolis-Hastings \citep{metropolis1953equation} or Gibbs sampler \citep{geman1984stochastic}, it provides robust performances and easily reaches convergence even for very complex models, e.g.: posterior density characterized by complex geometries, multi-layered hierarchical models with many parameters, models depending on large sets of latent variables, etc.
One of the advantages of \texttt{Stan} is that it allows for an easy implementation of the No U-Turn Sampler (NUTS, \cite{hoffman2014no}).  
NUTS proved to perform at least as efficiently as the standard HMC but, generally, does not require any tuning of the hyperparameters governing the proposals. Hence, it sensibly reduces the computational burden and averts any user intervention or wasteful runs.
Another advantage of the NUTS algorithm is that situations in which the sampling cannot thoroughly explore the whole posterior distribution are easily detectable. Indeed, when the approximation of the Hamiltonian dynamic fails to reach specific areas without departing from the original Hamiltonian value, the so-called \textit{divergent} transitions arise. For more details we refer to \cite{betancourt2016diagnosing}. The \texttt{Stan} interface reports divergences as warnings and provide ways to access which iterations encountered them. The \texttt{bayesplot} package \citep{gabry2019visualization} can be used to visualize them and locate the areas in which the exploration failed. If no divergences occur, we can be confident that the chain was able to explore the whole domain of interest of the log-posterior density.

After few warm-up iterations, during which the NUTS automatically adapts its future behaviour to the shape of the posterior density, chain convergence and desirable accuracy are usually reached even in few iterations ($\approx 10^3$). Nevertheless, doing more iterations does not harm and longer chains lead to more robust result. 
When the log-posterior density is computationally intensive to compute or the geometry of the posterior is particularly complex, the approximation of the Hamiltonian dynamics can be significantly slowed down and negatively impact on the total run-time. 
For instance, for the model in Sec.~\ref{sec:methodology}, there are $T$ spatial vectors of random effects that contribute to the overall density. The evaluation of the contribution of each of these requires the computation of the corresponding prior, which in turn involves the computation of inverse and determinant of the $G\times G$ matrix $Q(\alpha, \bW)$. It is clear how the naive implementation of such a model is all but efficient. Nevertheless, we can exploit two facts in order to ease computations. First, the spatial covariance structure does not vary along time, especially along iterations; this implies that we can compute the inverse and determinant only once in advance without repeating the same calculations repeatedly. Second, each region has only a few neighbours, and the matrix $Q(\alpha, \bW)$ is not full, paving the way for efficient algebraic solutions. 
In practice, many efficient strategies can be adopted in order to alleviate the computational burden by speeding up linear algebra operations. Here, we based ours on the \textit{Exact-sparse CAR} elaborated in \cite{joseph2016exact}, which accrues significant computational efficiency by more than halving the needed run-time\footnote{The computational gain is inversely proportional to the degree of the network defining neighbourhoods}. The original code has been slightly modified in order to include the temporal AR structure of our spatio-temporal CAR. The core of the \texttt{Stan} program needed to update the CAR-AR random effects is presented in Algorithm \ref{sparsecar}.
The full codes to reproduce the results presented in Sec.~\ref{sec:results} are available in a public \texttt{GitHub} repository accessible at \url{https://github.com/minmar94/Covid19-Spatial}.

\section{Application to COVID-19 incidence in Italy}
\label{sec:results}

\begin{figure}[ht]
    \centering
    \begin{subfigure}[b]{0.45\linewidth}
    \centering
    \includegraphics[width = \textwidth]{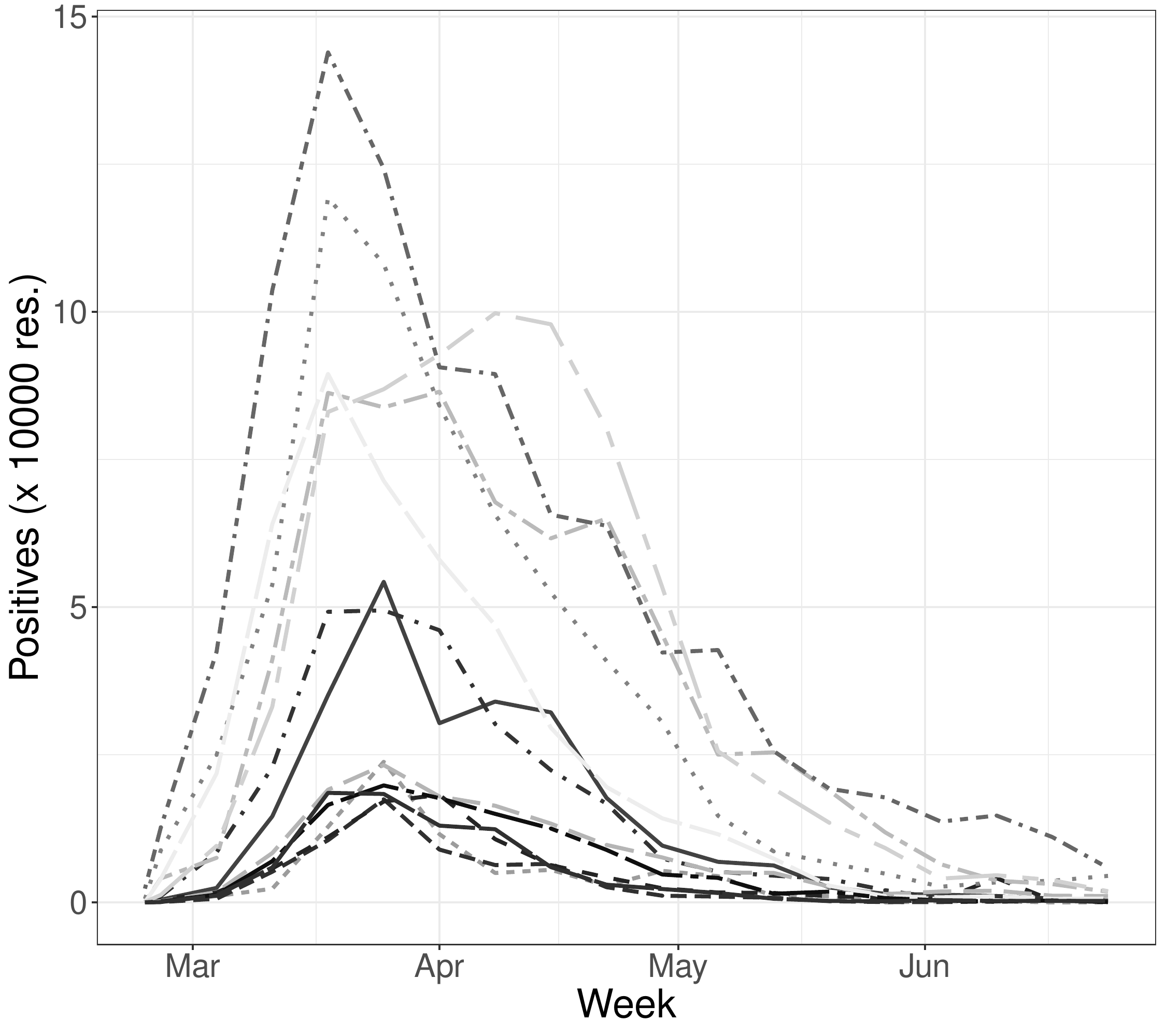}
    \caption{}
    \label{fig:WeeklyTsI}
    \end{subfigure}
    \hspace{0.05cm}
     \begin{subfigure}[b]{0.45\linewidth}
    \centering
    \includegraphics[width = \textwidth]{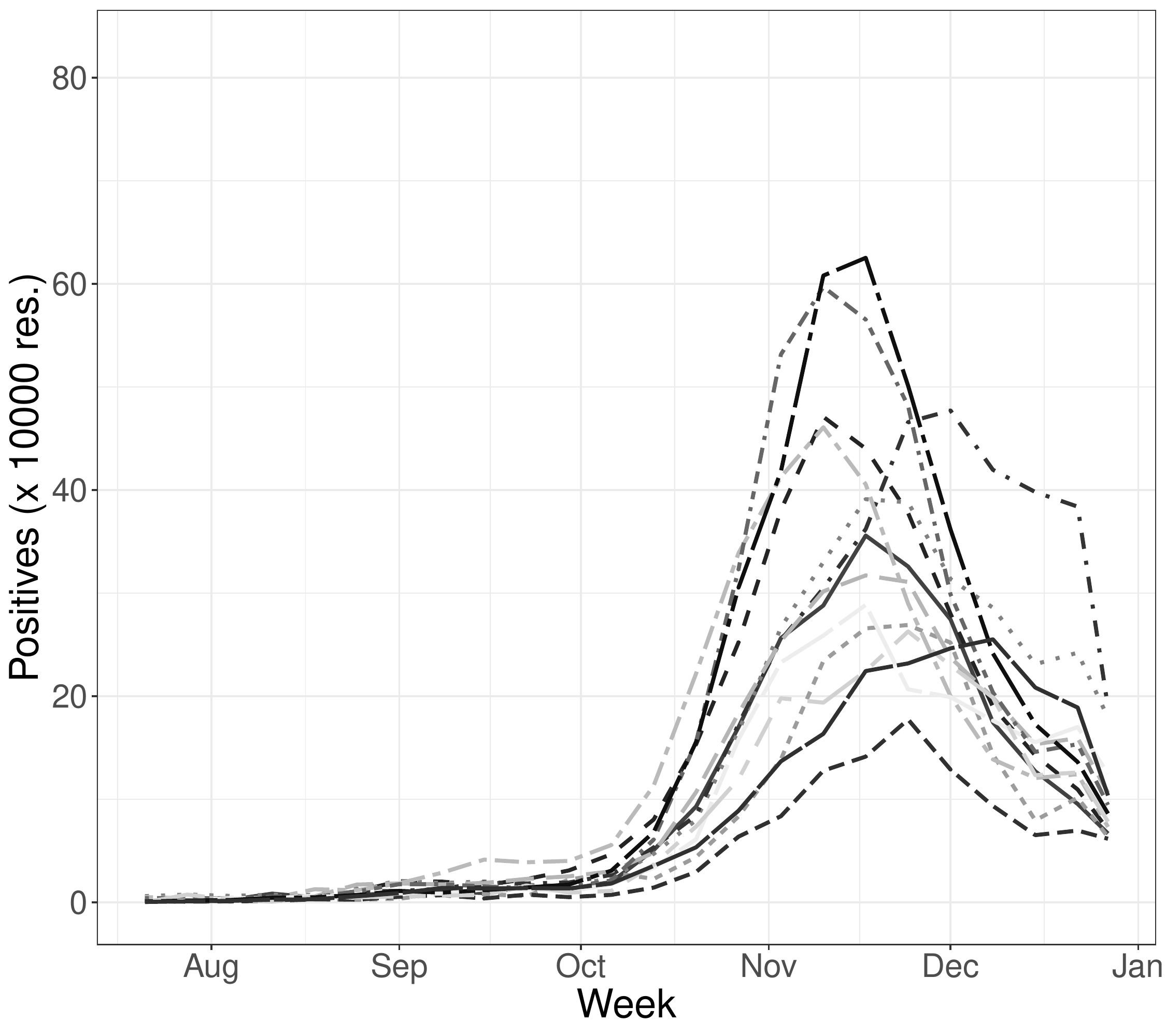}
    \caption{}
    \label{fig:WeeklyTsII}
    \end{subfigure}
    \caption{Regional weekly time series for \textit{positives} during the first (a) and second (c) wave. 
    }
\end{figure}

We test and compare our proposals defined in Sec.~\ref{sec:methodology} and \ref{sec:estimation} on the data described in Sec.~\ref{sec:avadata}.
We consider the time series of the \textit{weekly positives} at the regional ($G = 20$) level, for the first and the second wave of the epidemic.
From an epidemiological perspective, there is no strict definition for what is or is not an epidemic wave (or phase). However, the scientific community agrees on the fact that the word \textit{wave} implies a natural pattern of peaks and valleys, suggesting that even during a lull, future outbreaks of the disease are possible.
Our proposal is able to model only one epidemic wave at a time, since the Richards' curve entails a single peak time and height. 
The latter implies that the start and end date of each wave must be set by the researcher.
Albeit this is a drawback of our approach, it can be easily seen through sensitivity analyses that results do not drastically depend on this choice.
We mention here the work by \cite{bartolucci2021spatio}, which can flexibly model more than one wave at a time, but at the price of not being able to explicitly estimate important characteristics
of each wave (e.g., peak time, onset time, etc.). Similarly, \cite{farcomeni2020ensemble} does not require to identify a time frame for waves, but it is restricted to
short term predictions. 

In our application, we set the 24th of February, 2020, as the start date of the first wave, namely when systematic data recording started, while the 19th of July 2020 is set as the end date. That is the day in which discos and pubs were re-opened after the lockdown period (a total of 22 weeks). For the second wave, the 20th of July 2020 was set as the start date, while the 27th of December 2020 was set as the end date (for a total of 24 weeks), which corresponds to the end of the last week of the year and, more importantly, is the day the vaccine campaign began in all Europe (a.k.a. \textit{V-day}).
The regional time series of the \textit{weekly positives} for both waves are reported in Figure \ref{fig:WeeklyTsI} and Figure \ref{fig:WeeklyTsII}, respectively. It can be seen
that the second wave had a slower onset (due to the seasonality of infections in early Summer) but a much higher peak for most regions. That is not only due to a larger number of
infected with respect to the first wave (which has mostly hit only the northern part of Italy) but also to the much larger proportion of identified cases. 

We recall that we used the logarithm of the number of residents scaled by a factor of $10^5$ as an offset, essentially studying the number of positives each $10,000$ residents rather then crude incidence. This is necessary in order to be able to compare different regions, which can have very different number of infected only due to a very different number of residents. 
We also included the number of total weekly swabs (standardised) as covariate, to take into account different contact tracing efforts.
The number of positive swabs can be assumed to be negatively associated to the proportion of undetected cases, and is one of the official indicators of the World Health Organization. 

We also compared two different specifications for the adjacency matrix in the CAR-AR model. The first matrix, which we refer to as $\bW_1$, specifies a neighbourhood structure based on proximity flows and the availability of direct train, flights, and ferry connections.
This matrix has been estimated in \cite{della2020network}, and can lead to distant regions to be neighbours because of, for instance, frequent internal flight connections.
The original matrix is a weighted measure of commuters' flow and is not symmetric since exchanges may have different magnitudes in the two directions. As a fast and viable solution to symmetrise the matrix in this application, we decided to dichotomise it. We set $w^*_{ij} = 1$ if there exists a positive flow in at least one of the two directions.
The second adjacency matrix, which we refer to as $\bW_2$, is the most typically adopted network defined on regions' mutual geographical position. In our application, we considered a first-order structure, where only pairs of regions sharing at least one land border are considered as neighbours. 

\begin{figure}[ht]
    \centering
    \begin{subfigure}[b]{0.45\linewidth}
    \centering
    \includegraphics[width = 0.9\textwidth]{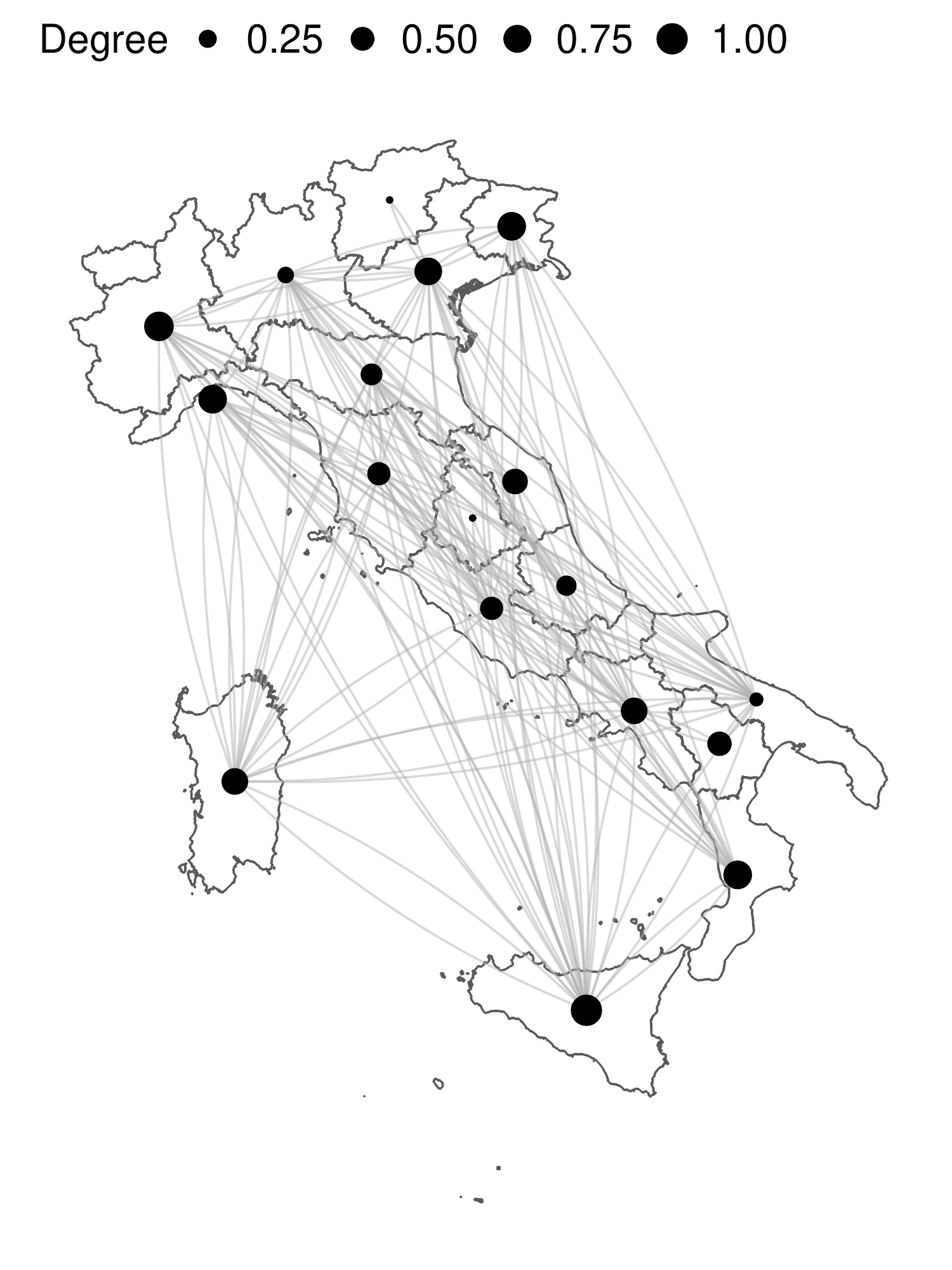}
    \caption{}
    \label{fig:Itgraph1}
    \end{subfigure}
    \hspace{0.05cm}
    \begin{subfigure}[b]{0.45\linewidth}
     \centering
    \includegraphics[width = 0.9\textwidth]{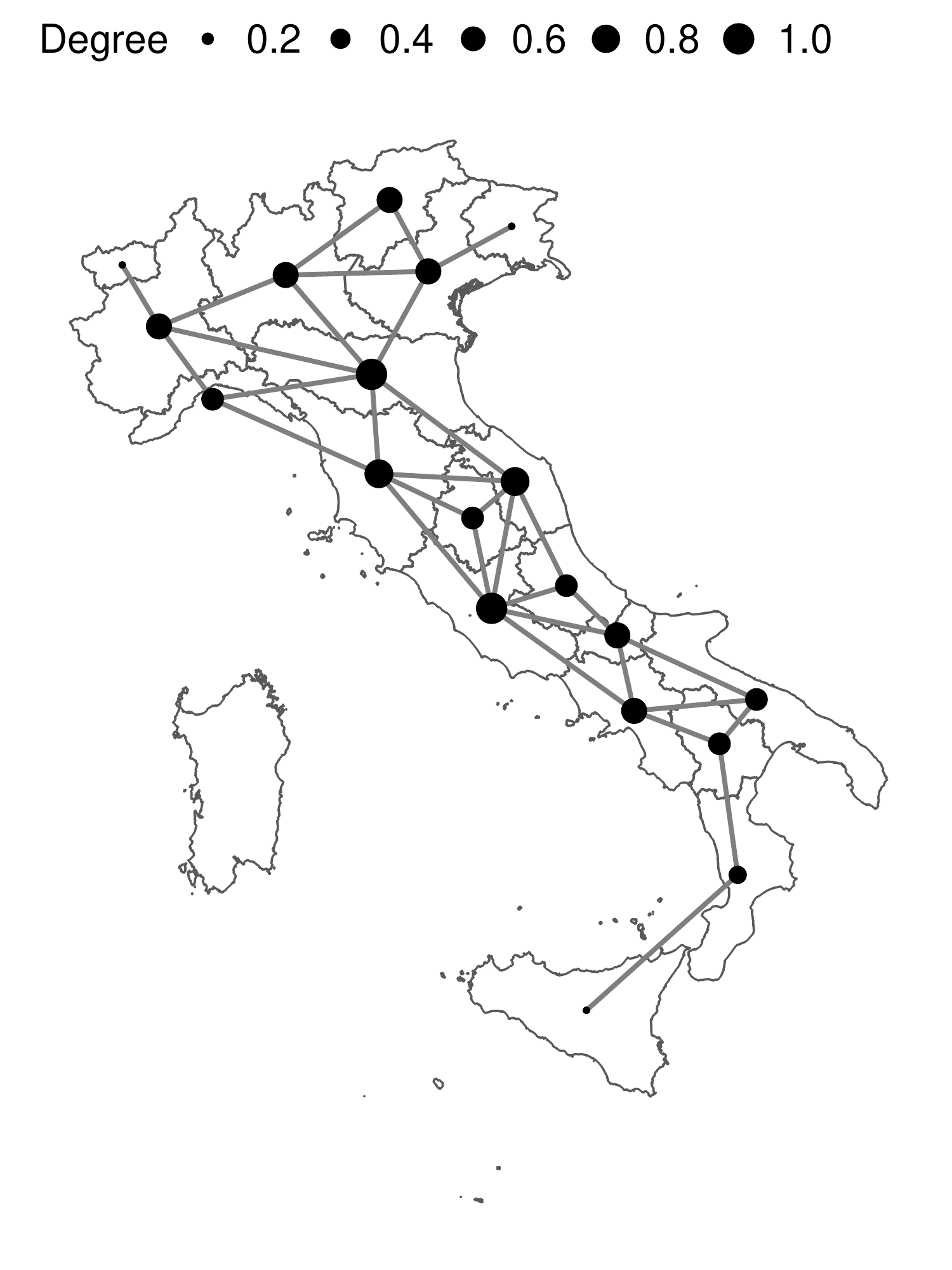}
    \caption{}
    \label{fig:Itgraph2}
    \end{subfigure}
    \caption{Network structure of the two adjacency matrices considered: $\bW_1$ (a) and $\bW_2$ (b).}
\end{figure}

The two different neighbourhood structures are shown in Fig.~\ref{fig:Itgraph1} and Fig.~\ref{fig:Itgraph2}, respectively. In particular, we report the number of edges (connections) and the (scaled) degree of each region. We notice that using $\bW_1$ we end up with 18 out of 20 regions that have at least one connection (Molise and Valle d'Aosta have none), three of them having 12 neighbours (which is the mode), and where Sicilia is the most connected area with 15 neighbours.
On the other hand, using $\bW_2$ we end up with seven regions that have 3 neighbours; two regions that have 6 neighbours, while Sardegna, which is an island, has no connections.
For Sicilia, which is also an island, we selected Calabria as the only neighbour. The two regions are separated by very few kilometres of sea (the Strait of Messina), with extremely frequent ferry connections. 

As a baseline model for comparison, we also considered the possibility of a completely disconnected graph $\bW_0=\lsq 0\rsq_{ij},\;\forall\,i,j$, hence assuming complete spatial independence between regions. Nevertheless, being temporal dependence undeniably present in the observed series, we always retain the temporal AR structure between subsequent vectors $\bphi_{t-1},\bphi_t,\;t=2,\dots,T$. 
As a matter of fact, preliminary runs that neglected this feature of the date produced way worse results (especially in terms of out-of-sample performances) that will not be reported in the sequel.

\subsection{Main results}
\label{subsec:results}
For the sake of brevity, we will refer to the model ignoring spatial dependence with the fully disconnected graph $\bW_0$ as $M_0$, and as $M_1$ and $M_2$ to the models including spatial dependence using $\bW_1$ and $\bW_2$ as adjacency matrices, respectively. We considered these three dependence structures for the model with one common Richards' curve, which we name \textit{common}, and the model with region-specific Richards' curves, which we name \textit{regional}.

For all models considered, we ran two separate chains for $10,000$ iterations, allowing \texttt{Stan} to perform $5,000$ warm-up iterations each, which were discarded for inferential purposes.

We computed several metrics in order to compare the goodness-of-fit and predictive performance of the model's alternatives. The large flexibility of the space-time random effects specification easily makes the model fit the observed set of data almost perfectly. This feature exposes the typical in-sample metrics to over-fitting, flawing any sensible interpretation of the results, and would inevitably favour the highly parametrized \textit{regional} model. 
Therefore, we decided to avert the over-fitting issues by artificially subtracting $15\%$ randomly selected points from each region's time series. 
These are treated as missing data in the estimation process, and the ability to reconstruct the missing pieces properly is then verified in terms of various metrics: \textit{Coverage},  \textit{Root Mean Squared Error} (RMSE) and \textit{Predictive Interval Width} (PIW).
Comparison of these three metrics for the three dependence structures, with common and regional Richards, for the two waves, are presented in Table \ref{tab:comVsReg}.

\begin{table}[ht]
\centering
{\fontsize{9}{9}\selectfont
    \begin{tabular}{cc|ccc|ccc}
      & & \multicolumn{3}{c}{Common} &  \multicolumn{3}{|c}{Regional} \\
     \cline{3-8}
     \textbf{Wave} & \textbf{Metric} & $M_0$ & $M_1$ & $M_2$ & $M_0$ & $M_1$ & $M_2$\\
     \toprule
     \multirow{3}{*}{I} & Coverage & 0.98 & 0.98 & 0.98 & 0.96 & 1 & 0.96 \\
      & PIW & 1535 & 1178 & 1144 & 3311 & 846 & 1017 \\
       & RMSE & 423 & 184 & 272 & 399 & 331 & 314\\
       \midrule
       \multirow{3}{*}{II} & Coverage & 0.96 & 0.97 & 0.92 & 0.97 & 0.96 & 0.92 \\
      & PIW & 33393 & 4497 & 4046 & 16900 & 3131 & 3121\\
       & RMSE & 12841 & 910 & 995 & 3669 & 1008 & 1038 \\
       \bottomrule
\end{tabular}
}
\caption{Out-of-sample predictive performances of our proposals with a common or region-specific Richards' curve in the first and the second wave.}
\label{tab:comVsReg}
\end{table}

We can clearly observe how the out-of-sample performances are comparable across the \textit{common} and \textit{regional} specifications. The coverage is close (actually larger in most cases) to the $95\%$ nominal level in both cases, for all the dependence structures. 
$M_1$ and $M_2$, under both the common and regional specification of the logistic trend, show equivalent coverage and similar PIWs. 
However, the common specification provides more accurate out-of-sample predictions in terms of RMSE, whenever the random-effects account for the spatial dependence ($M_1$ and $M_2$). 
Therefore, given the comparable out-of-sample performances and the more parsimonious specification of the \textit{common} model, we chose this one as the preferred option. All future results will then be referred to this specification.

Parameter estimates for the spatial ($\alpha$) and temporal ($\rho$) auto-correlation, together with the swabs' effect ($\beta$) are reported in Table \ref{tab:sptempbetaparsest}. Here, we want to first highlight that there is a clear evidence of strong dependence both spatial and temporal, with values of $\hat{\rho}>0.8$ in all models for both waves. 
It is instead notable how the transport based graph detects low spatial dependence ($\hat{\alpha}\approx 0.14$) during the first wave, and a large spatial dependence during the second wave ($\hat{\alpha}\approx 0.93$). This change in the spatial correlation parameter between the first and the second wave for $M_1$, highlights the different type of non-pharmaceutical measures that were adopted to contain the spread of the contagion and the estimated values are completely reasonable, given the harder block to inter-regional movements that characterized the first wave as compared to more liberal mobility policies that accompanied the second one. On the contrary, the geographic vicinity effect is stable across the two waves, probably capturing similarities between close regions that depend on shared unobserved characteristics more than on people exchange. 

The parameter $\beta$ represents the effect of additional swabs on the number of detected positives. Being the swabs variable standardized, this does not allow for a trivial interpretation. However, we can observe a positive effect which was more evident during the first wave than the second wave. This happens unsurprisingly, since the testing efforts were not yet at full capacity during the first outbreak, with many undetected cases, detected as soon as additional testing hubs were made available.

\begin{table}[t]
\renewcommand{\arraystretch}{0.8}
    \centering
    {\fontsize{9}{9}\selectfont
    \begin{tabular}{c|c|ccc}
    \toprule
         \textbf{Wave} & \textbf{Param.} & $M_0$ & $M_1$ & $M_2$\\
         \midrule
         \multirow{3}{*}{\textbf{I}} & $\alpha$ & -- & 0.14 (0.02,\,0.21) & 0.76 (0.71,\,0.81) \\
         & $\rho$ & 0.89 (0.87,\,0.91) & 0.88 (0.90,\,0.93) & 0.86 (0.85,\,0.89) \\
         & $\beta$ & 0.36 (0.26,\,0.44) &  0.34 (0.25,\,0.42) & 0.21 (0.14,\,0.29) \\
         \midrule
         \multirow{3}{*}{\textbf{II}} & $\alpha$ & -- & 0.93 (0.92,\,0.95) & 0.87 (0.85,\,0.90) \\
         & $\rho$ & 0.88 (0.86,\,0.90) & 0.87 (0.85,\,0.89) & 0.82 (0.80,\,0.85) \\
         & $\beta$ & 0.42 (0.38,\,0.46) & 0.27 (0.24,\,0.30) & 0.13 (0.09,\,0.16) \\
         \bottomrule
    \end{tabular}
    }
    \caption{Comparison of parameters' estimates for the spatial ($\alpha$) and temporal ($\rho$) auto-correlation, and for the swabs' effect in the first and the second wave.}
    \label{tab:sptempbetaparsest}
\end{table}

\begin{table}[ht]
\setlength{\tabcolsep}{3pt}
\renewcommand{\arraystretch}{0.8}
    \centering
    {\fontsize{9}{9}\selectfont
    \begin{tabular}{l|c|ccccc}
    \toprule
         \textbf{Wave} & \textbf{Model} & $b$ & $r$ & $h$ & $p$ & $s$ \\ 
         \midrule
         \multirow{3}{*}{\textbf{I}} & $M_0$ & \small 0.05 (0.04,\,0.06) & \small 23 (20,\,27) & \small 0.62 (0.60,\,0.64) & \small 2.0 (1.5,\,2.5) & \small 7.8 (6.3,\,9.9) \\
         & $M_1$ &
            \small 0.06 (0.05,\,0.07) & \small 20 (17,\,22) & \small 0.62 (0.59,\,0.65) & \small 2.2 (1.7,\,2.8) & \small 7.9 (5.5,\,9.3) \\
         & $M_2$ &
            \small 0.05 (0.04,\,0.06) & \small 26 (21,\,31) & \small 0.61 (0.58,\,0.65) & \small 2.2 (1.5,\,2.9) & \small 7.8 (5.2,\,9.3) \\
            \midrule
          \multirow{3}{*}{\textbf{II}} & $M_0$ &
            \small $7\cdot 10^{-5}$ ($1\cdot 10^{-6}$,\,$1\cdot 10^{-3}$) & \small 158 (143,\,172) & \small 3.46 (3.26,\,3.63) & \small 23.2 (23.1,\,23.3) & \small 0.06 (0.05,0.07) \\
         & $M_1$ &
            \small $2\cdot 10^{-4}$ ($3\cdot 10^{-5}$,\,$7\cdot 10^{-3}$) & \small 178 (127,\,215) & \small 2.72 (2.33,\,3.08) & \small 22.9 (22.8,\,23.2) & \small 0.09 (0.07,0.10) \\
        & $M_2$ &
            \small $4\cdot10^{-4}$ ($3\cdot10^{-6}$,\,$1\cdot10^{-2}$) & \small 194 (163,\,220) & \small 3.50 (3.20,\,3.70) & \small 23.1 (22.9,\,23.2) & \small 0.06 (0.05,\,0.07) \\
        \bottomrule 
    \end{tabular}
    }
    \caption{Parameters' estimates of the Richards' curve for the first and the second wave.}
    \label{tab:richparsest}
\end{table}
Table \ref{tab:richparsest} shows the estimated parameters of the common Richards in all settings. We here recall that $b$ represents the baseline (endemic rate), $r$ the final size of the outbreak (in terms of cases every $10,000$ residents), $h$ the contagion speed, $p$ the lag-phase and $s$ the asymmetry.
We can clearly observe how the second wave is characterized by a larger final outbreak size, a larger endemic rate and a longer lag-phase (meaning the curve approximates exponential growth for a longer time window) in all cases. Furthermore, while the first outbreak was characterized by positive asymmetric behaviour (with a fast and sudden growth followed by a long descending phase) the second wave presented a negative asymmetric evolution, probably because of the softer lockdown measures undertaken.
Indeed, the positive asymmetry characterising the first wave reflects the hard containment measures implemented by the Italian government at the beginning of the epidemic (March 2020), which were gradually loosened. On the contrary, the second wave experienced a negative asymmetry as the prevention policies were mild at the beginning of the second outbreak (mid Summer 2020) and were suddenly strengthened following the abrupt increase of positive cases in late November 2020.

\begin{figure}[t]
    \centering
    \begin{subfigure}[b]{0.31\linewidth}
        \includegraphics[width = \textwidth]{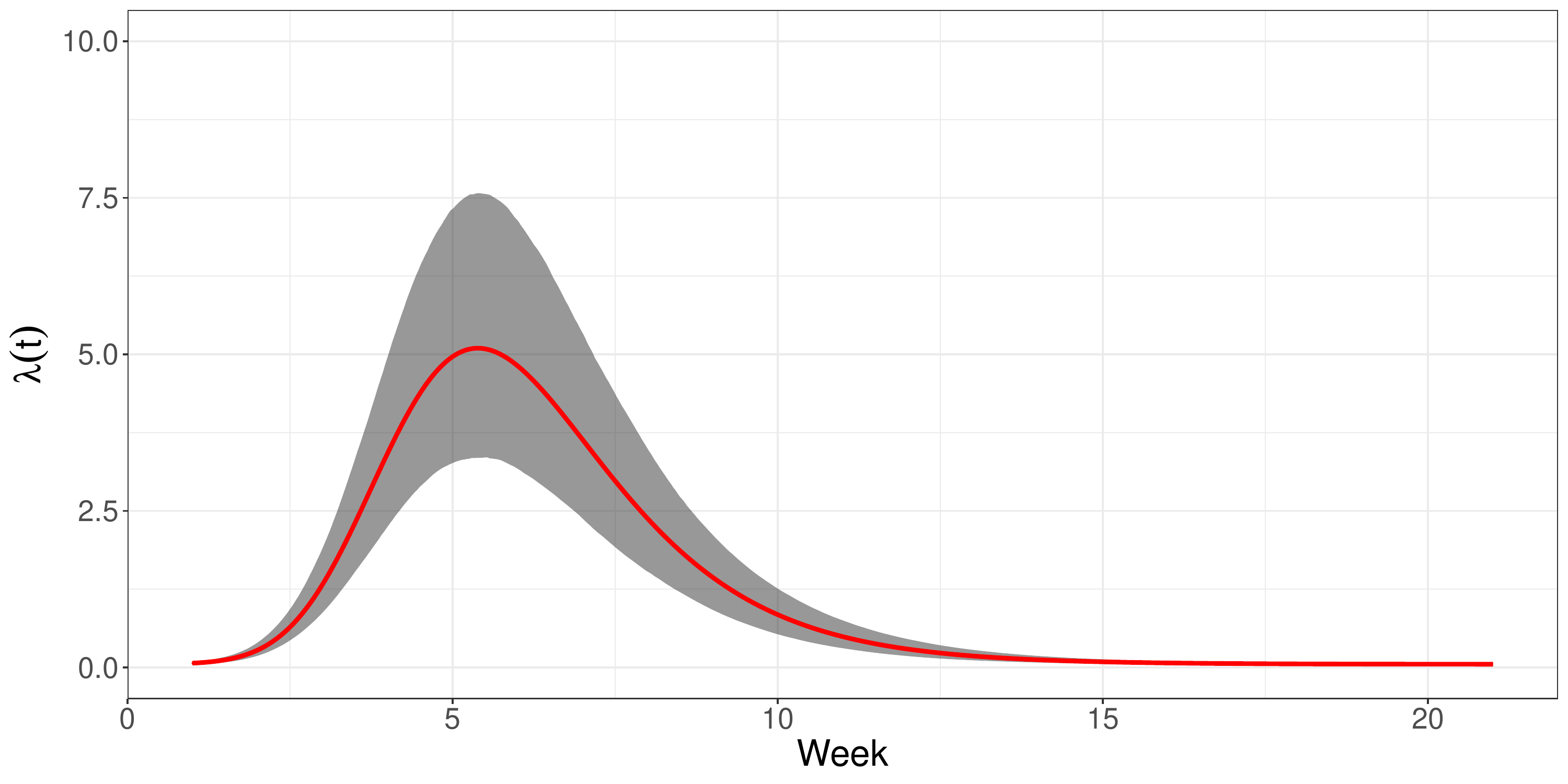}
    \caption{$M_0$}
    \label{CRIndIwave}
    \end{subfigure}
    \begin{subfigure}[b]{0.31\linewidth}
        \includegraphics[width = \textwidth]{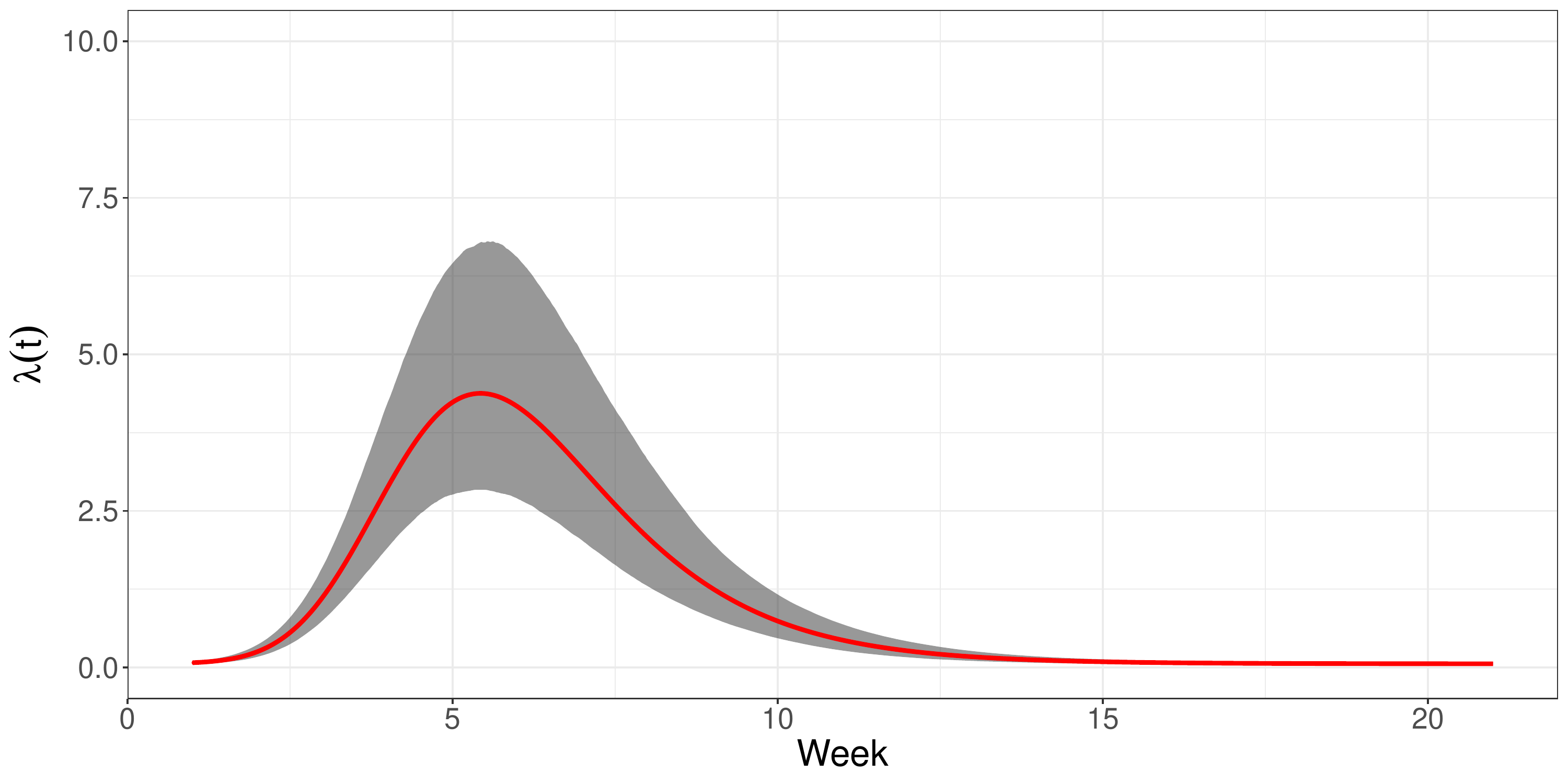}
    \caption{$M_1$}
    \label{CRTrIwave}
    \end{subfigure}
     \begin{subfigure}[b]{0.31\linewidth}
        \includegraphics[width = \textwidth]{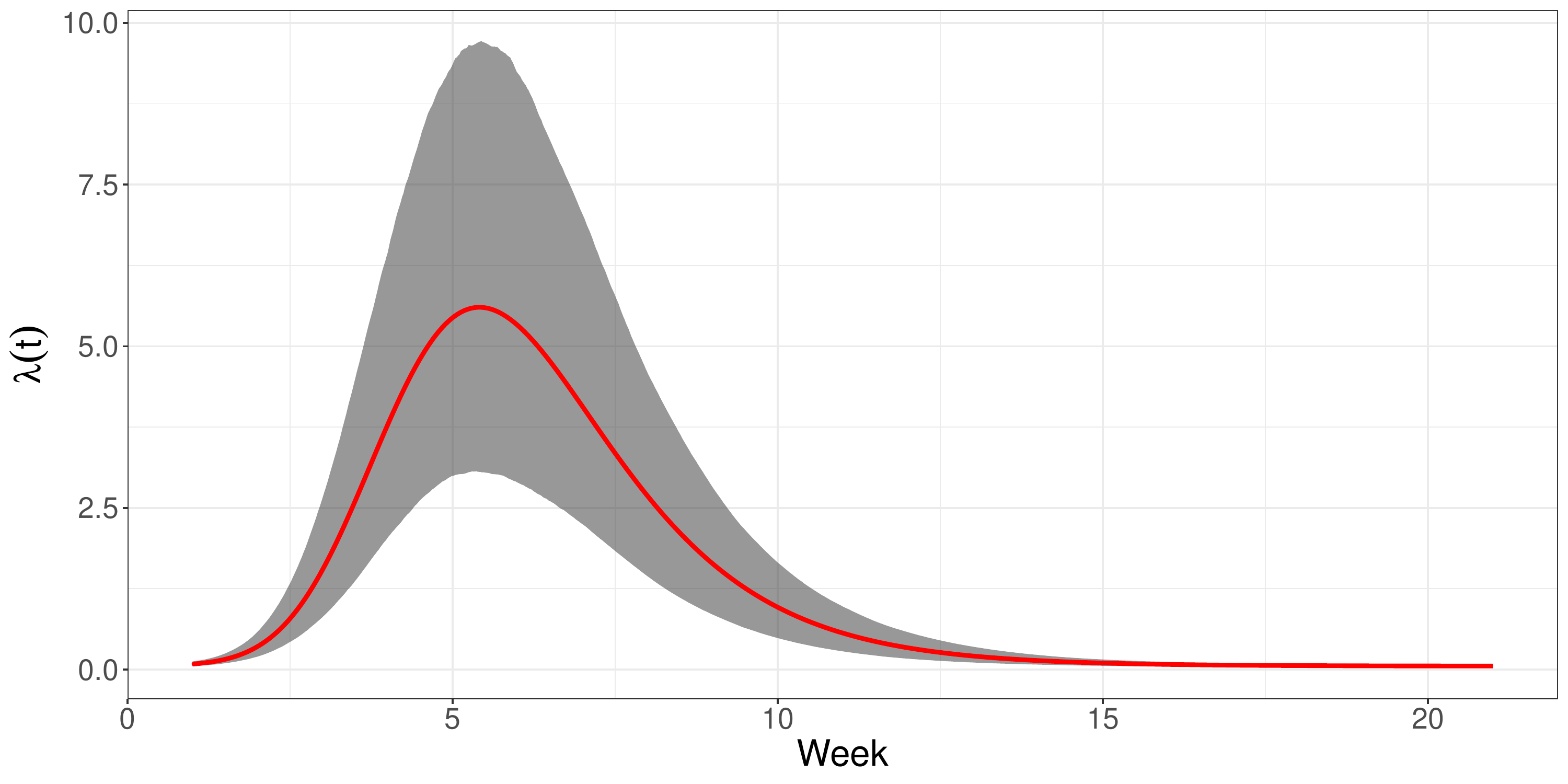}
    \caption{$M_2$}
    \label{CRGeoIwave}
    \end{subfigure}
    \begin{subfigure}[b]{0.31\linewidth}
        \includegraphics[width = \textwidth]{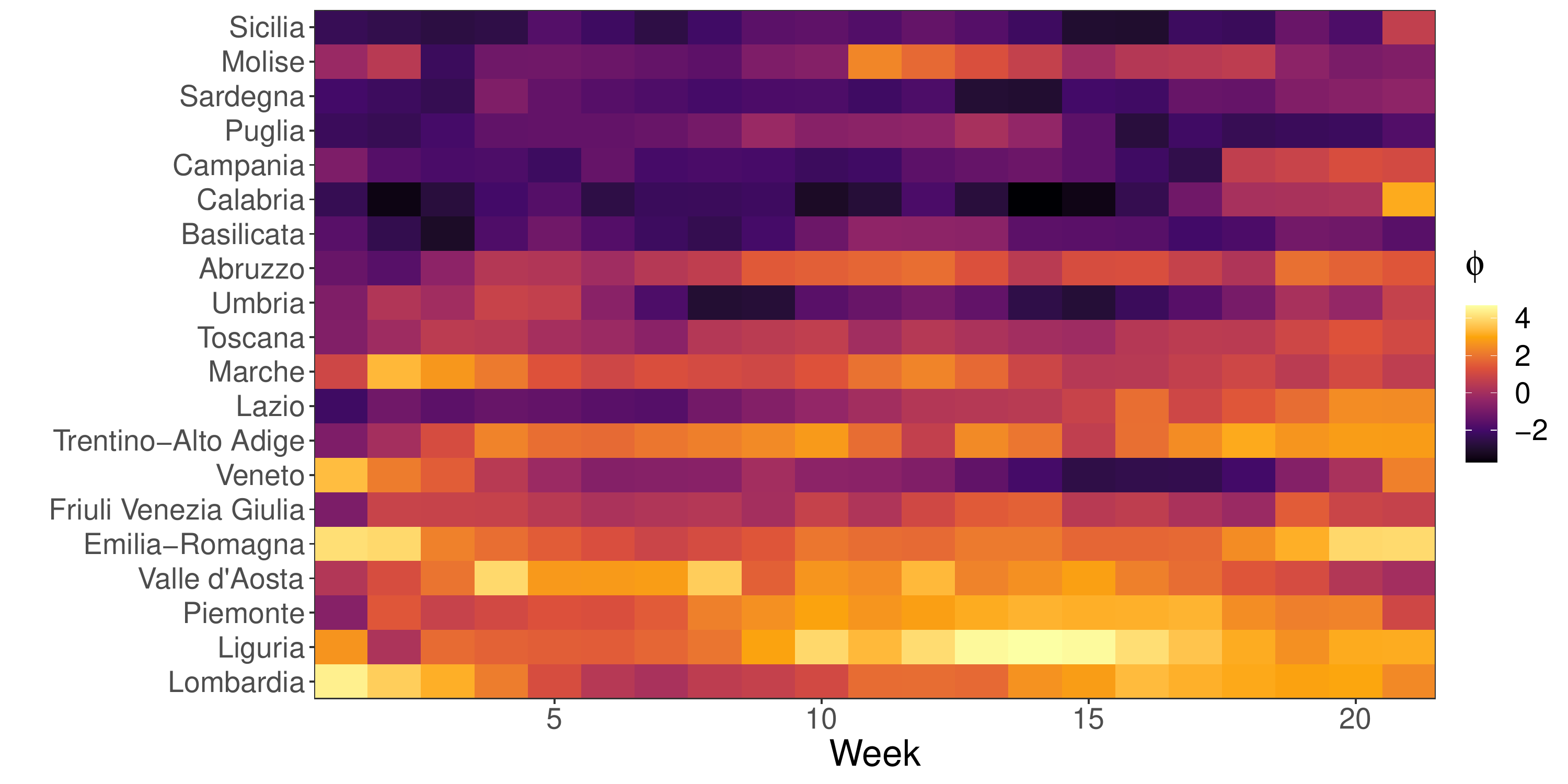}
    \caption{$M_0$}
    \label{PhiCRIndIwave}
    \end{subfigure}
    \begin{subfigure}[b]{0.31\linewidth}
        \includegraphics[width = \textwidth]{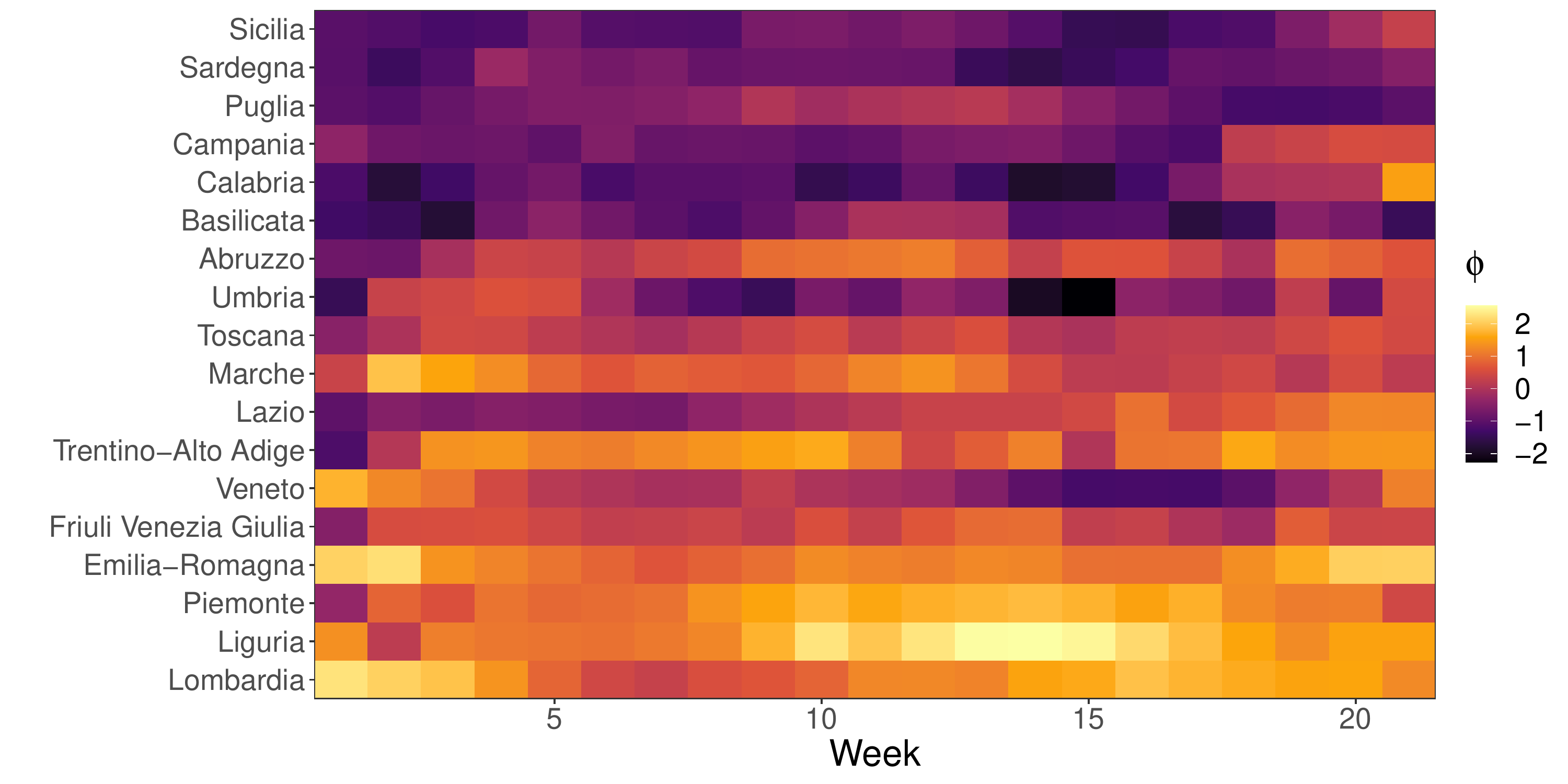}
    \caption{$M_1$}
    \label{PhiCRTrIwave}
    \end{subfigure}
    \begin{subfigure}[b]{0.31\linewidth}
        \includegraphics[width = \textwidth]{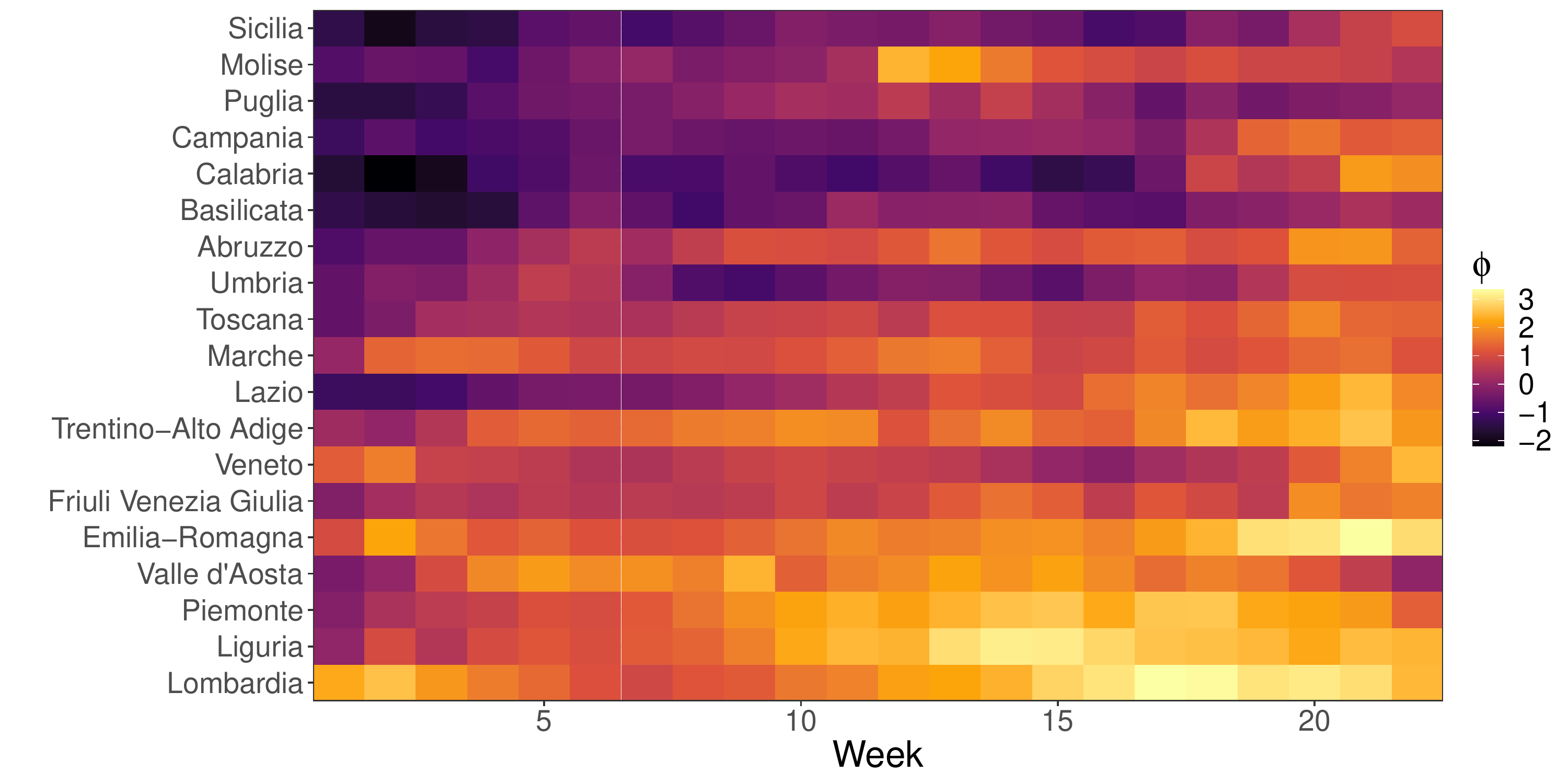}
    \caption{$M_2$}
    \label{PhiCRGeoIwave}
    \end{subfigure}
    \caption{Common Richards' curve for the first wave for the different specifications of the random effect (top panels); Posterior mean of the random-effect (bottom panels).}
\end{figure}

\begin{figure}[t]
    \centering
    \begin{subfigure}[b]{0.31\linewidth}
        \includegraphics[width = \textwidth]{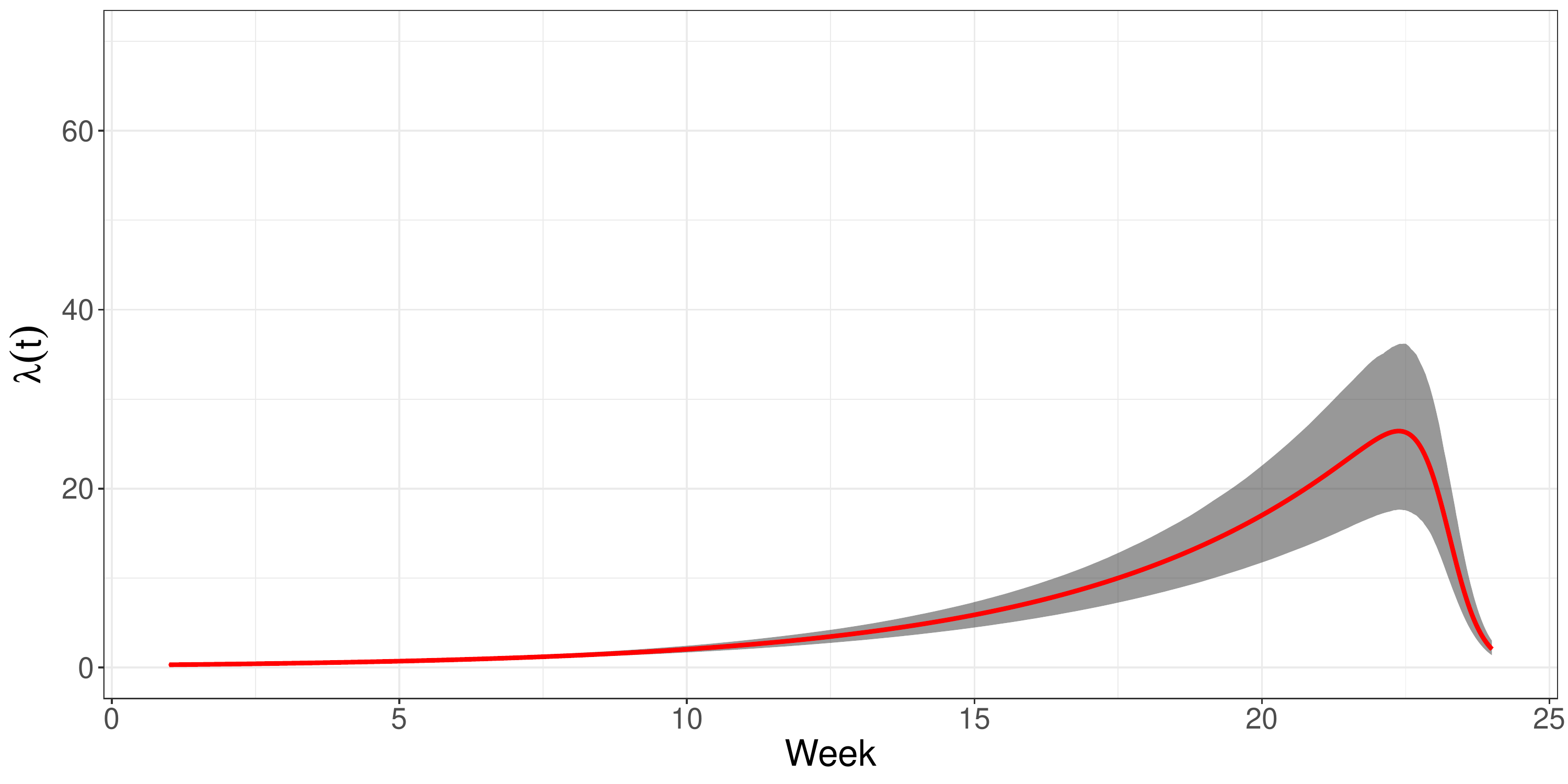}
    \caption{$M_0$}
    \label{CRIndIIwave}
    \end{subfigure}
    \begin{subfigure}[b]{0.31\linewidth}
        \includegraphics[width = \textwidth]{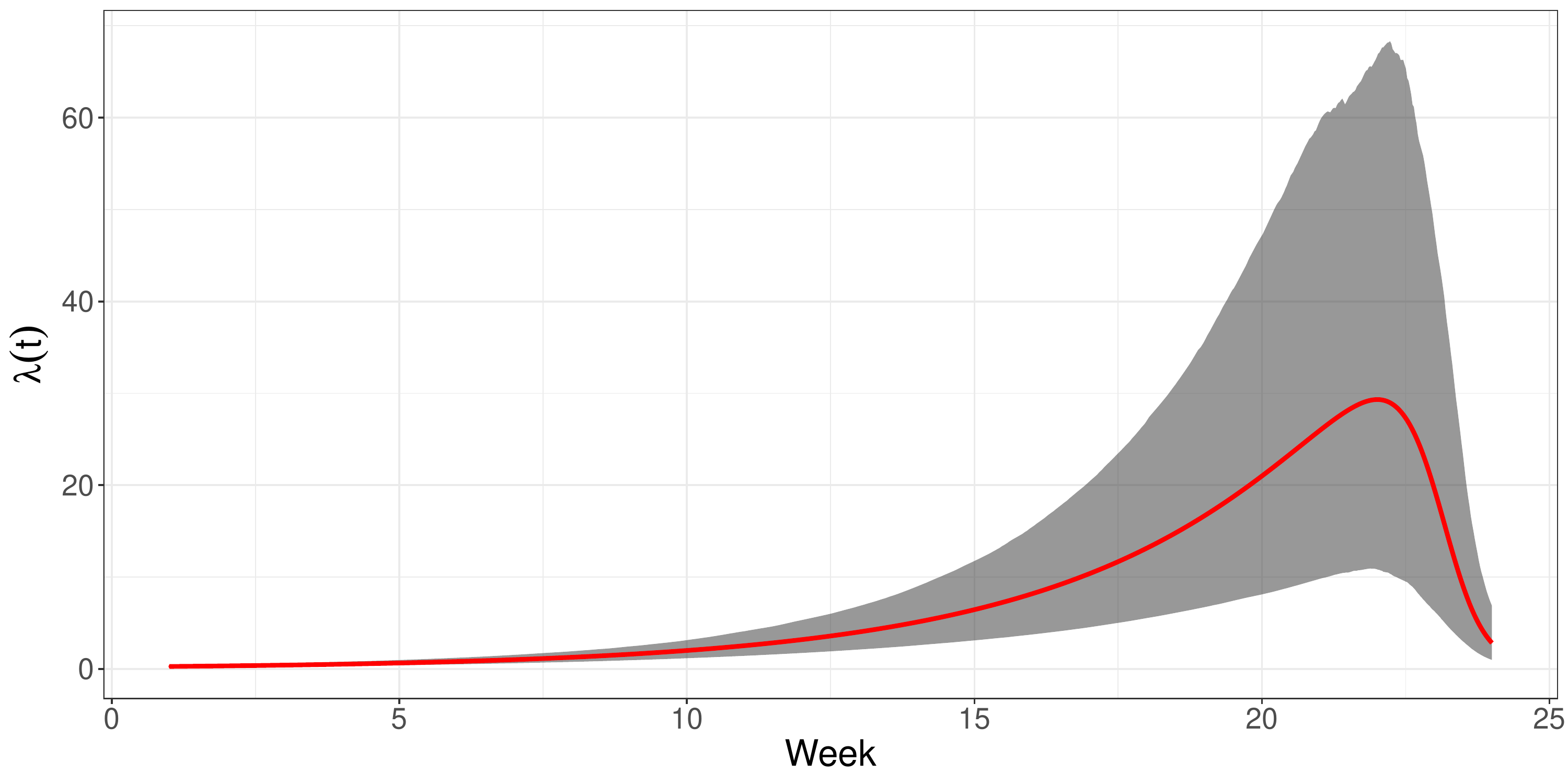}
    \caption{$M_1$}
     \label{CRTrIIwave}
    \end{subfigure}
    \begin{subfigure}[b]{0.31\linewidth}
        \includegraphics[width = \textwidth]{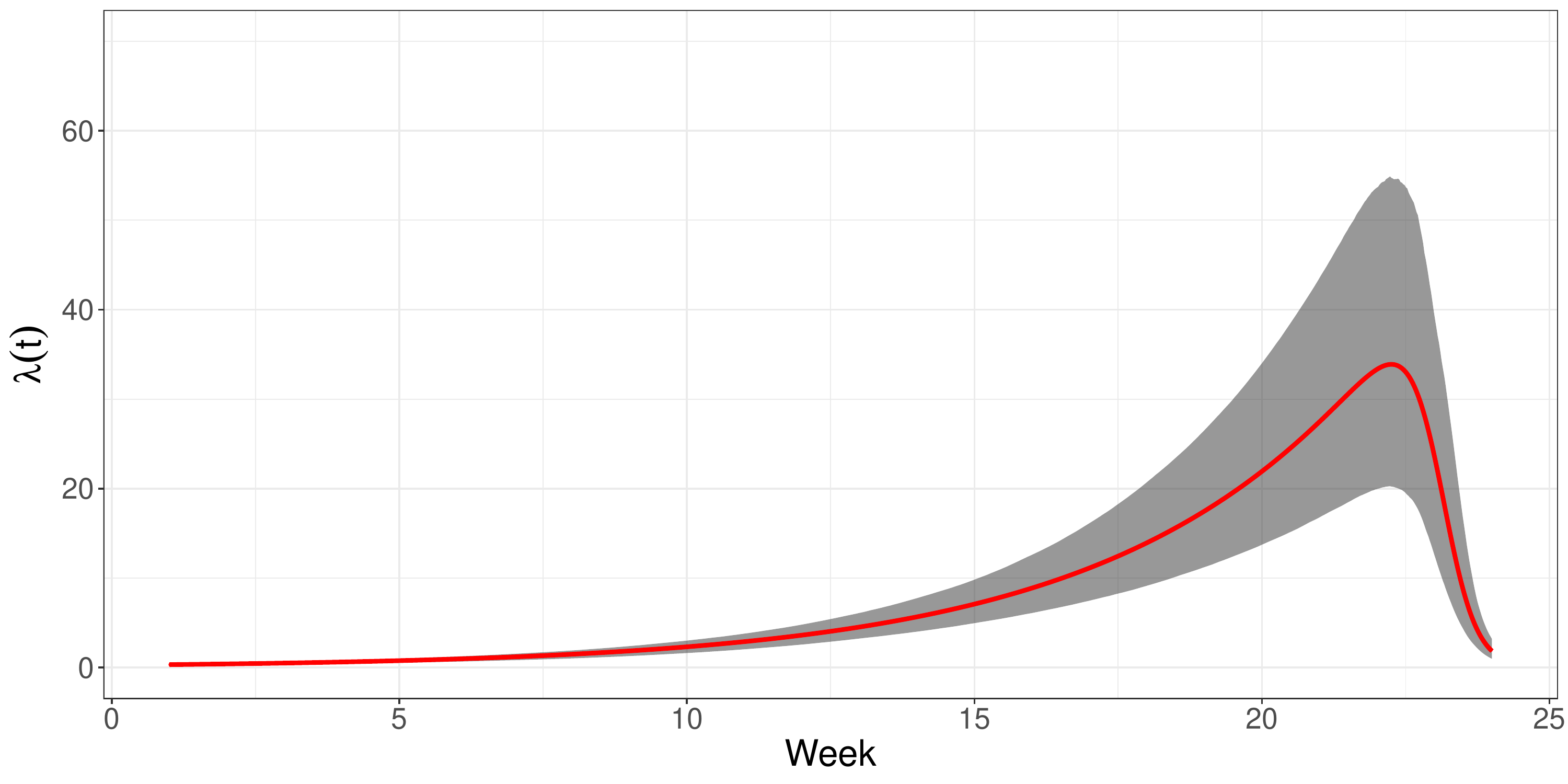}
    \caption{$M_2$}
     \label{CRGeoIIwave}
    \end{subfigure}
    \begin{subfigure}[b]{0.31\linewidth}
        \includegraphics[width = \textwidth]{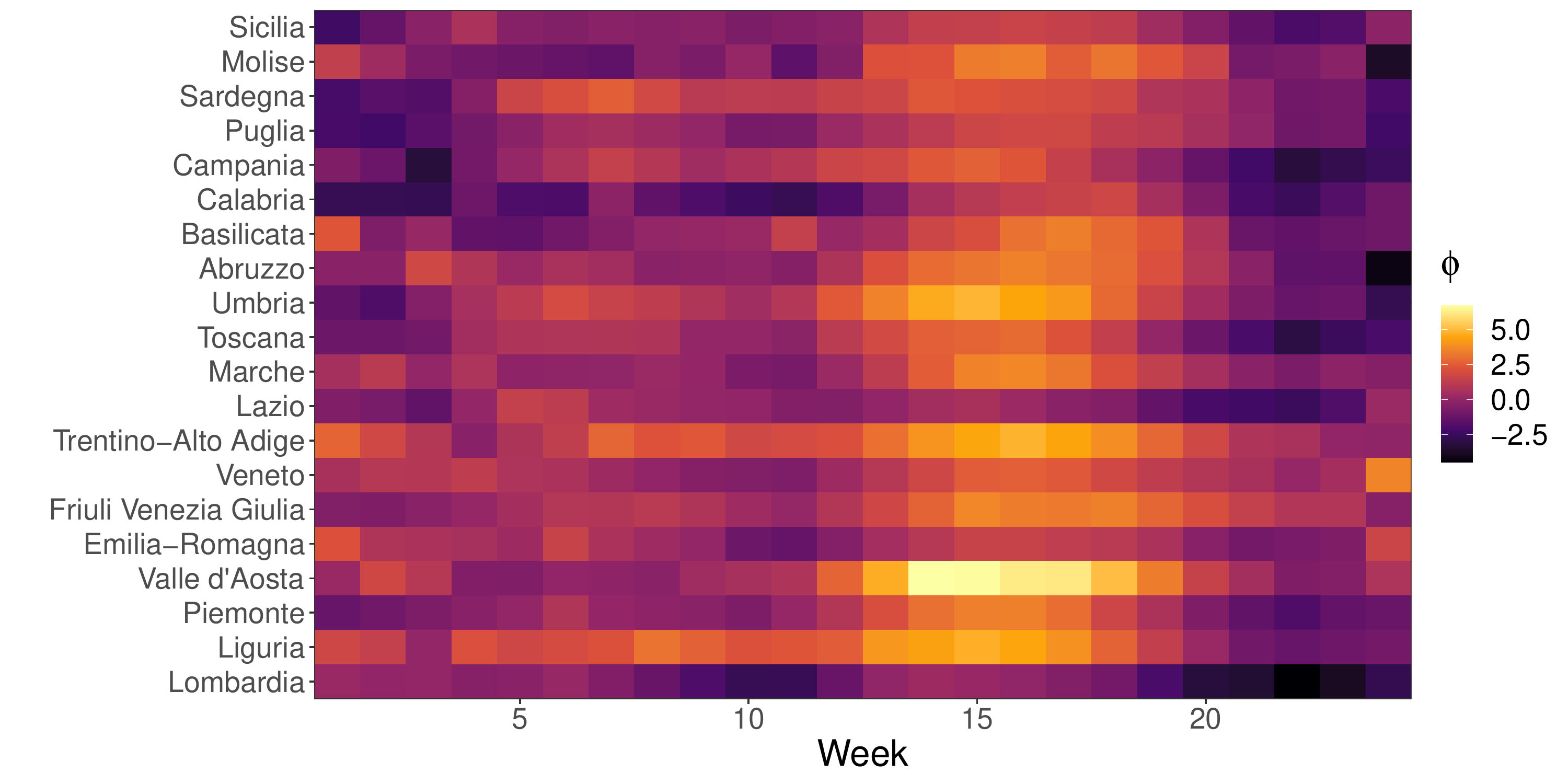}
    \caption{$M_0$}
    \label{PhiCRIndIIwave}
    \end{subfigure}
    \begin{subfigure}[b]{0.31\linewidth}
        \includegraphics[width = \textwidth]{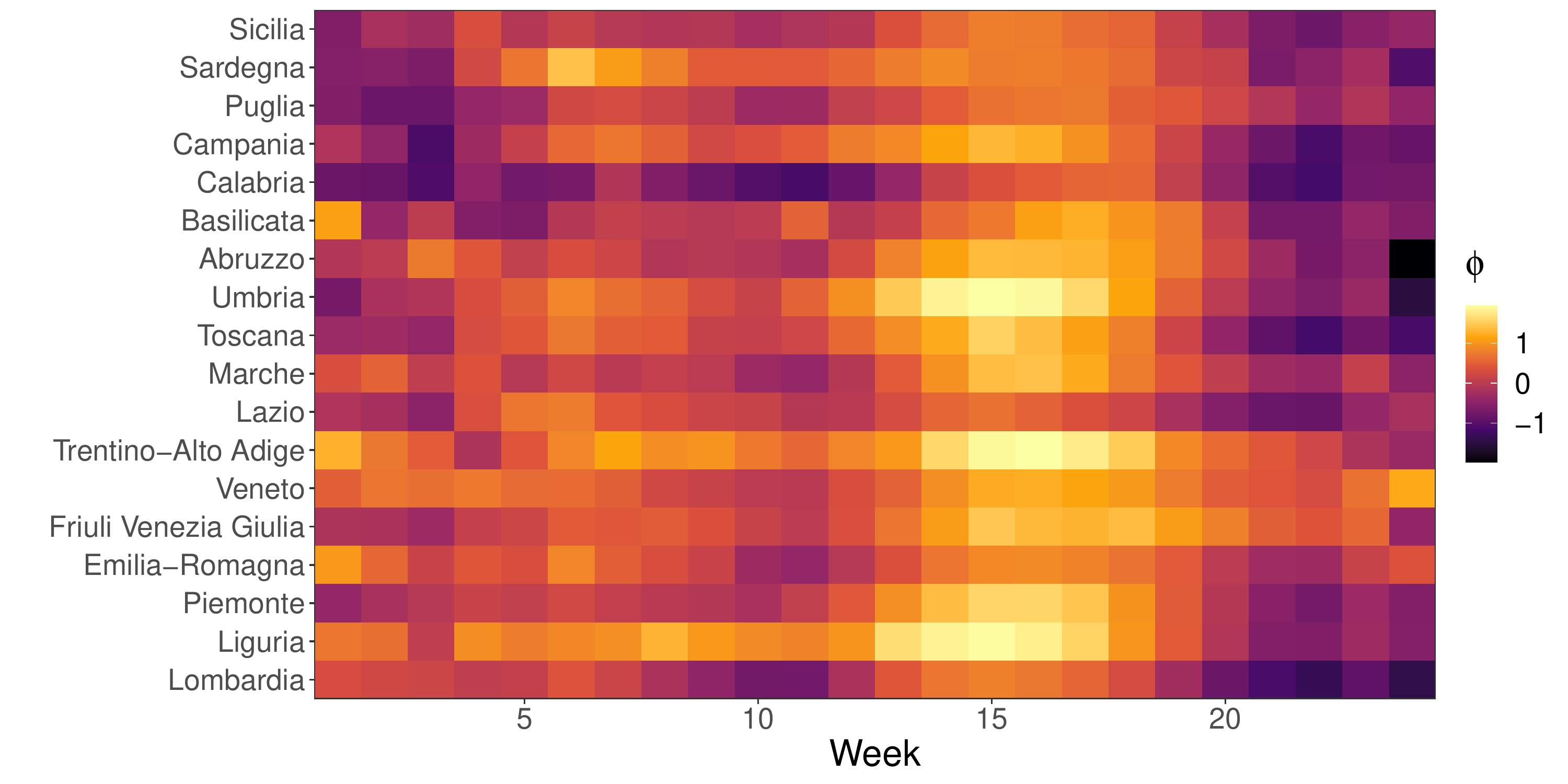}
    \caption{$M_1$}
    \label{PhiCRTrIIwave}
    \end{subfigure}
    \begin{subfigure}[b]{0.31\linewidth}
        \includegraphics[width = \textwidth]{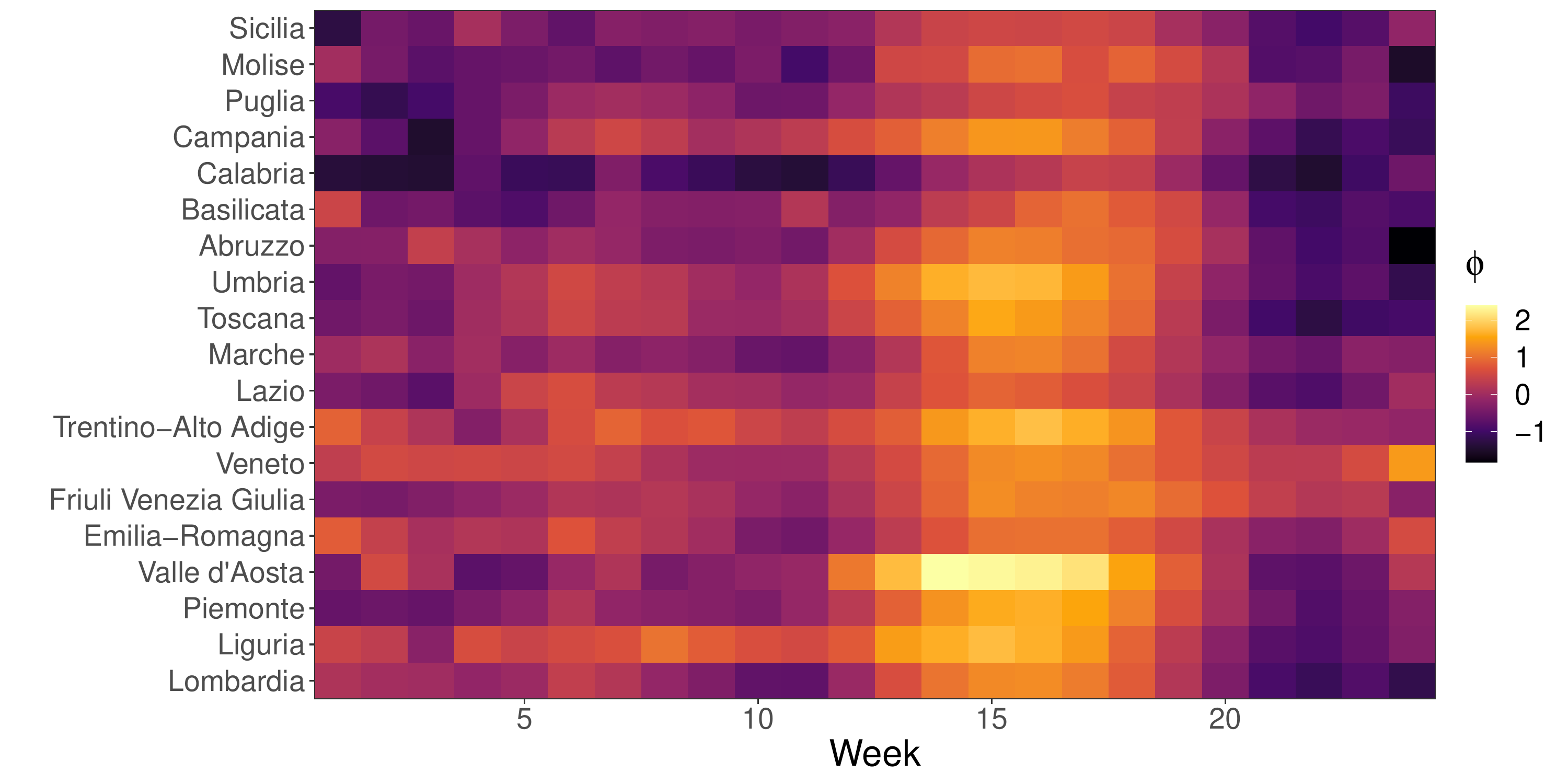}
    \caption{$M_2$}
    \label{PhiCRGeoIIwave}
    \end{subfigure}
    \caption{Common Richards' curve for the second wave for the different specifications of the random effect (top panels); Posterior mean of the random-effect (bottom panels).}
\end{figure}

Fig.~\ref{CRIndIwave}-\ref{CRGeoIwave} and Fig.~\ref{CRIndIIwave}-\ref{CRGeoIIwave} show the estimated common Richards' curves (red solid line) by the proposed models with the associated uncertainty (grey areas represent the $95\%$ credible intervals) for the first and the second wave, respectively. In Fig.~\ref{PhiCRIndIwave}-\ref{PhiCRGeoIwave} and Fig.~\ref{PhiCRIndIIwave}-\ref{PhiCRGeoIIwave} we instead report the heatmaps of the estimated spatio-temporal effect for each model specification for the first and the second wave, respectively.
The estimated common Richards' curve and random effects during the first wave highlight how  deviations from the global average presented a strong geographic \textit{clustering effect}, as the number of positive cases increased from   the South to the North of Italy. On the contrary, there is relative homogeneity in the deviations of each region from the national epidemic at each time point during the second wave. This means that all regions experienced a similar epidemic trend in terms of shape but different in terms of relative magnitude. Notably, some peculiar regional behaviours are highlighted very clearly. For example, a sudden surge in the contagion between October and November 2020 experienced by Trentino Alto Adige and Umbria.
In general, we notice that a larger uncertainty characterizes the common Richards' estimated by $M_1$ for the second wave compared to the other two models. Considering that the PIWs (see Table \ref{tab:valmetrics}) do not vary much across the proposed dependence structures, this implies a stronger identification of the random effects, i.e. less variability of the random effects. We can then assume that this model provides a better description of the regional heterogeneity in the data.

In order to fully compare the different dependence structures on the space-time random effects for the \textit{common} model, we also evaluated the overall fitting performances in terms of two wide-scope indicators: the \textit{Watanabe-Akaike Information Criterion} (WAIC) \citep{watanabe2010asymptotic}, and the \textit{Leave-One-Out} (LOO) score as in \citep{vehtari2017practical}. These two metrics shall be considered as a proxy of the out-of-sample prediction accuracy, but can be directly computed from the fitted Bayesian model by retaining the log-likelihood values at the different steps of the chain \citep{vehtari2017practical}. 
Results of the estimated validation metrics for the three models for both the first and the second wave are reported in Table \ref{tab:valmetrics}. We notice that results are comparable, with $M_1$ performing slightly better in terms of RMSE, WAIC and LOO for both waves, guaranteeing a greater or equal coverage in both scenarios. Furthermore, comparing the in-sample and out-of-sample RMSE, we can see how the independent $M_0$ model strongly overfits on the training set. On the other hand, limiting and driving the behavior of the random effects through a spatial structure (such as $M_0$ and $M_1$) strongly improves the predictive power of the model and leads to way more reliable results.
All things considered, $M_1$ is then chosen as the best dependence structure, also in light of the appealing interpretation of the varying spatial dependence strength as expressed by $\alpha$ in the two waves (see Table \ref{tab:sptempbetaparsest}). Hence, the following results will be referred to this model. 

\begin{figure}[t]
    \centering
    \begin{subfigure}[b]{0.45\linewidth}
    \centering
    \includegraphics[width = \textwidth]{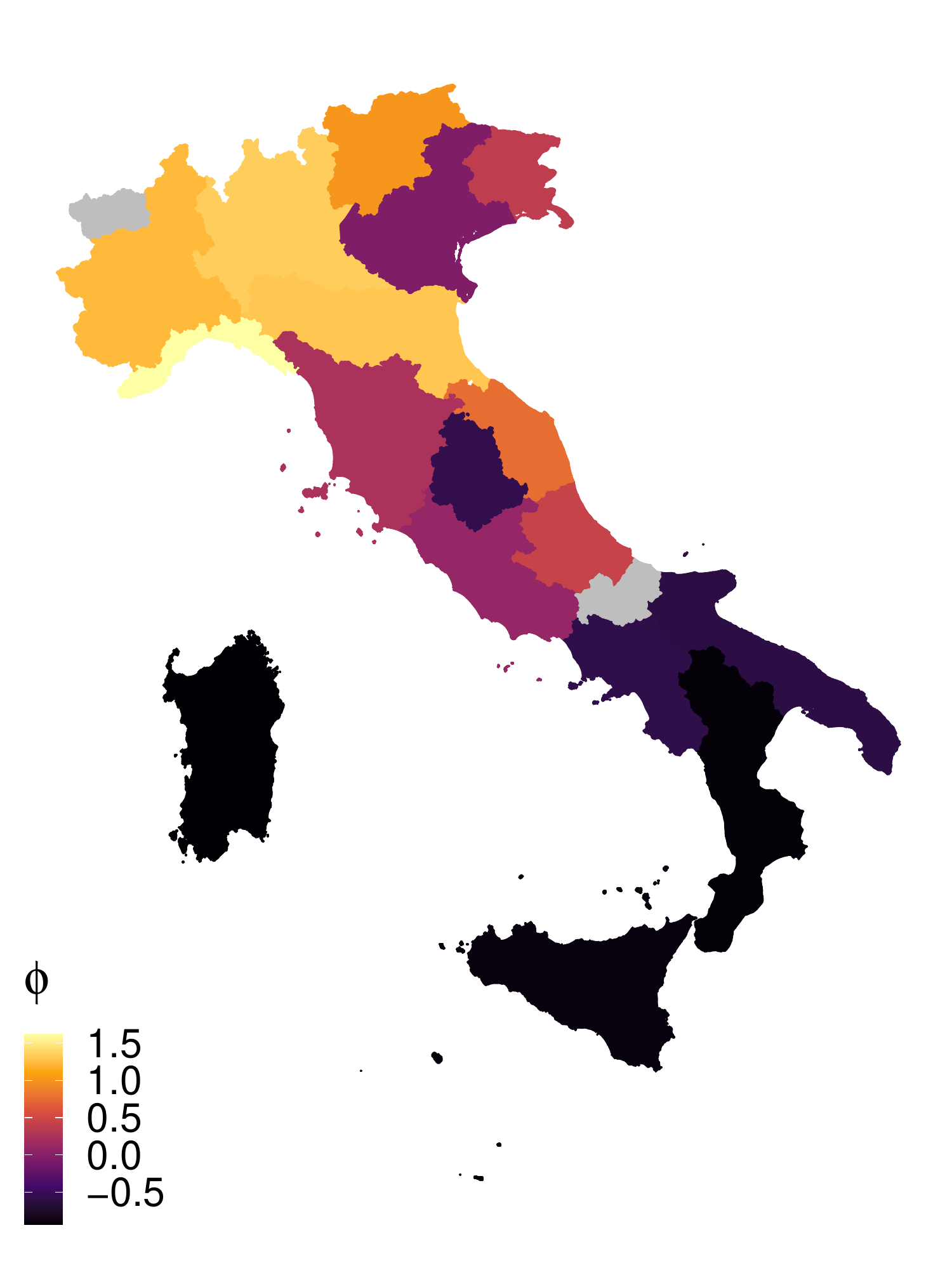}
    \caption{}
    \label{fig:PhiFlux}
    \end{subfigure}
    \hspace{0.05cm}
    \begin{subfigure}[b]{0.45\linewidth}
    \centering
    \includegraphics[width = \textwidth]{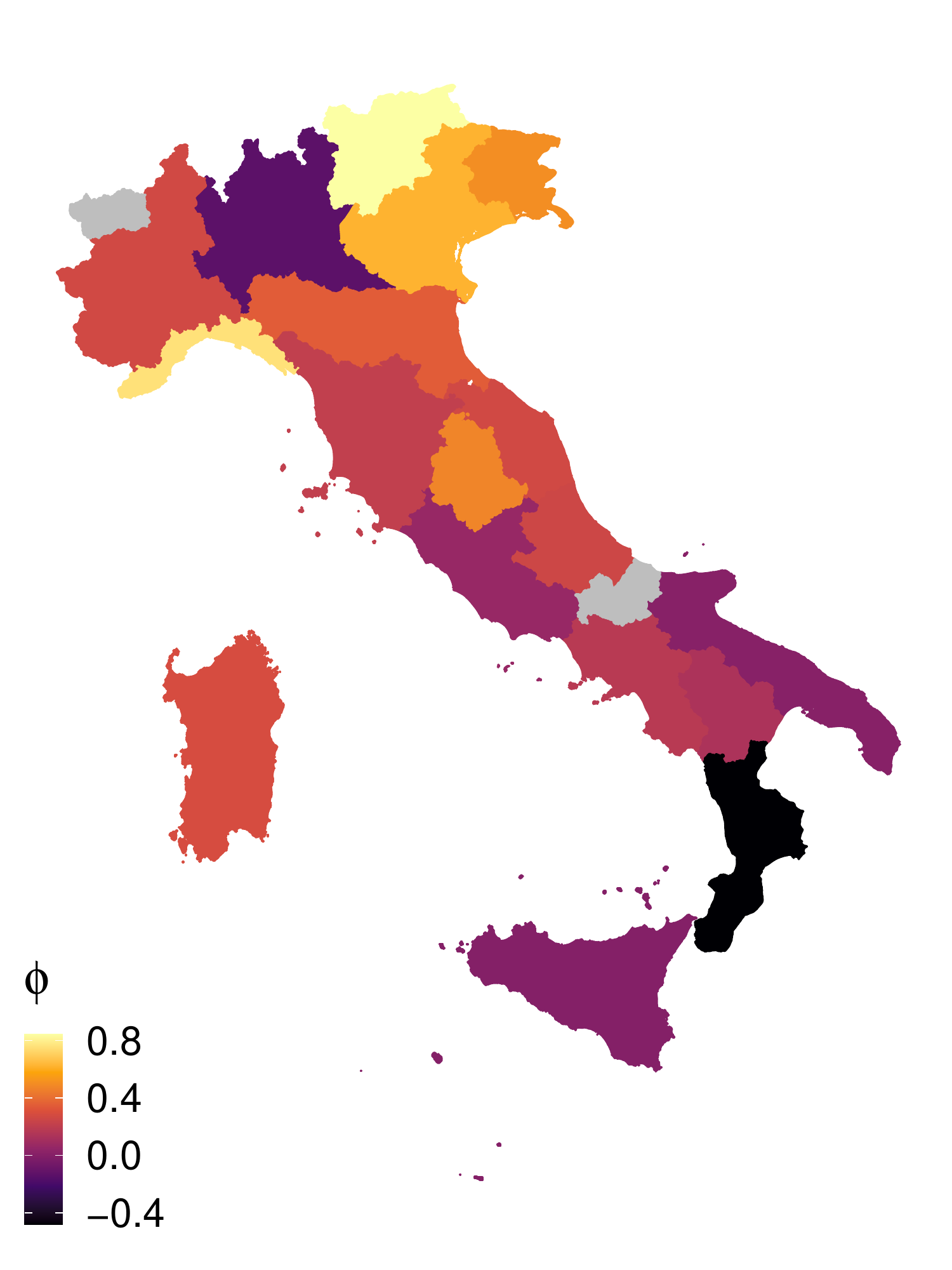}
    \caption{}
    \label{fig:PhiFluxII}
    \end{subfigure}
    \caption{Average estimated spatial random effect $\bar{\phi}_g$ by $M_1$ for the first (a) and the second (b) wave.}
\end{figure}

Figures~\ref{fig:PhiFlux}-\ref{fig:PhiFluxII} show the map of the temporal averages of the space time effects of each region: $\bar{\phi}_g=\sum_{t=1}^T\hat{\phi}_{gt}/T$.
These values can be interpreted as the effect of the over-dispersion on the contagion's spread due to the interactions with the neighbourhood and the auto-regressive term. That allows to verify what regions generally presented an infection rate larger than the national average along the two waves. As already pointed out, we can notice a stronger geographical clustering during the first wave, which is even more evident looking at the maps than at the heatmaps. Given the low value of the estimated $\alpha$ in the first wave, this effect is not really linked to the superimposed networking structure, but is an inherent characteristic of the data at regional level: the pandemic initially hit stronger the North of Italy and only later slowly spread to the South. 
On the contrary, the geographic clustering effect vanishes during the second wave and the coloring of the map looks smoother. Regions similarity is actually explained by people exchange and transportation between regions (larger value of $\alpha$). 
It is crucial to notice that, differently from what was often reported by the news in Italy, Lombardia did not perform worse in terms of positive cases with respect to the the rest of the country along the second wave (net of the tracking effort and regional offset). 
We added the maps obtained using $M_2$ for comparison in Fig. \ref{fig:PhiGeo} and \ref{fig:PhiGeoII} of the appendix, and we only report here that differences were negligible.

\begin{table}[ht]
\renewcommand{\arraystretch}{0.8}
    \centering
    {\fontsize{9}{9}\selectfont
    \begin{tabular}{c|c|cccc}
         \toprule
 \textbf{Wave} & \textbf{Model} & \textbf{Coverage} & \textbf{RMSE} & \textbf{WAIC} & \textbf{LOO} \\ 
 \midrule
     \multirow{3}{*}{\textbf{I}} & $M_0$ & 0.98 (1) & 423 (2.1) & 2869 & 3087 \\ 
& $M_1$ & 0.98 (1) & 184 (2.3) & 2650 & 2849 \\ 
& $M_2$ & 0.98 (0.99) & 272 (2.5) & 2774 & 2982 \\ 
\midrule
\multirow{3}{*}{\textbf{II}} & $M_0$ & 0.96 (0.99) & 12841 (2.8) & 4112 & 4393 \\ 
& $M_1$ & 0.97 (0.99) & 910 (4.6) & 3820 & 4080 \\
& $M_2$ & 0.92 (0.99) & 995 (4.1) & 3971 & 4252 \\ 
\bottomrule
    \end{tabular}
    }
    \caption{Validation metrics for the estimated models for the first and the second wave: coverage and rmse out-of-sample (in-sample), WAIC and LOO.}
    \label{tab:valmetrics}
\end{table}

\begin{figure}[H]
    \centering
    \begin{subfigure}[b]{0.22\linewidth}
        \includegraphics[width = \textwidth]{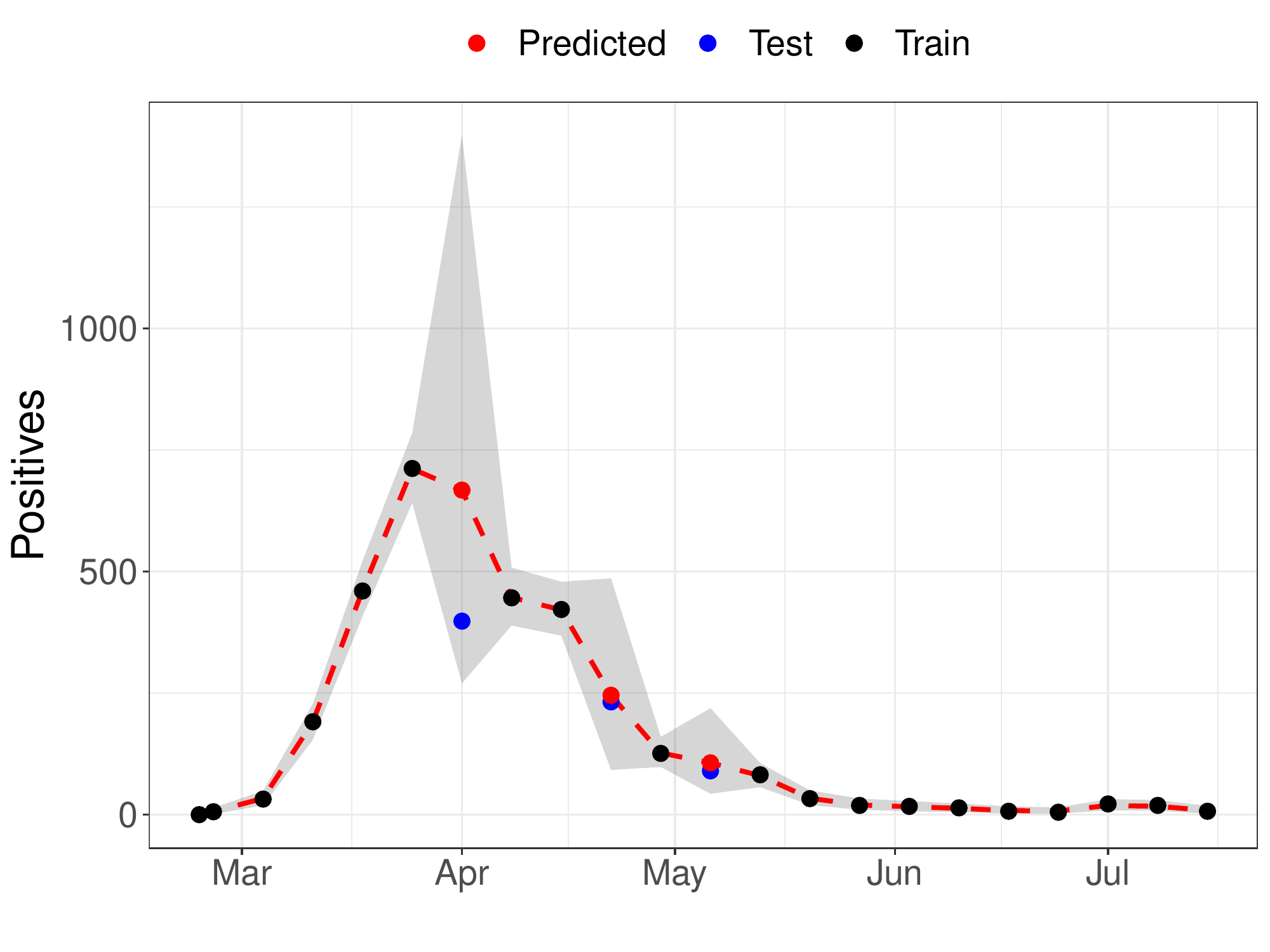}
    \caption{Abruzzo}
    \label{fig:moliseI}
    \end{subfigure}
    \begin{subfigure}[b]{0.22\linewidth}
        \includegraphics[width = \textwidth]{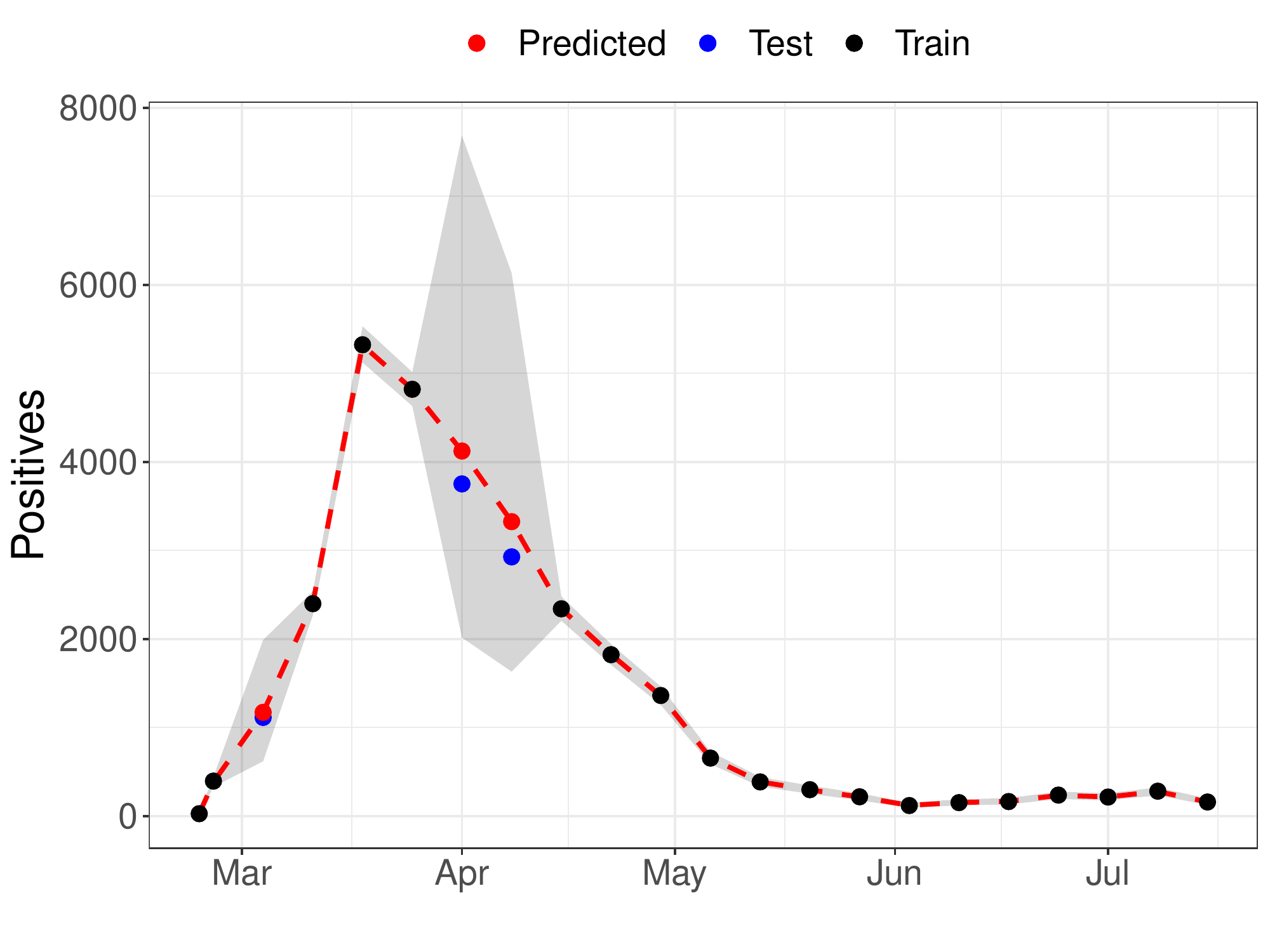}
    \caption{Emilia Romagna}
    \label{fig:lazioI}
    \end{subfigure}
    \begin{subfigure}[b]{0.22\linewidth}
        \includegraphics[width = \textwidth]{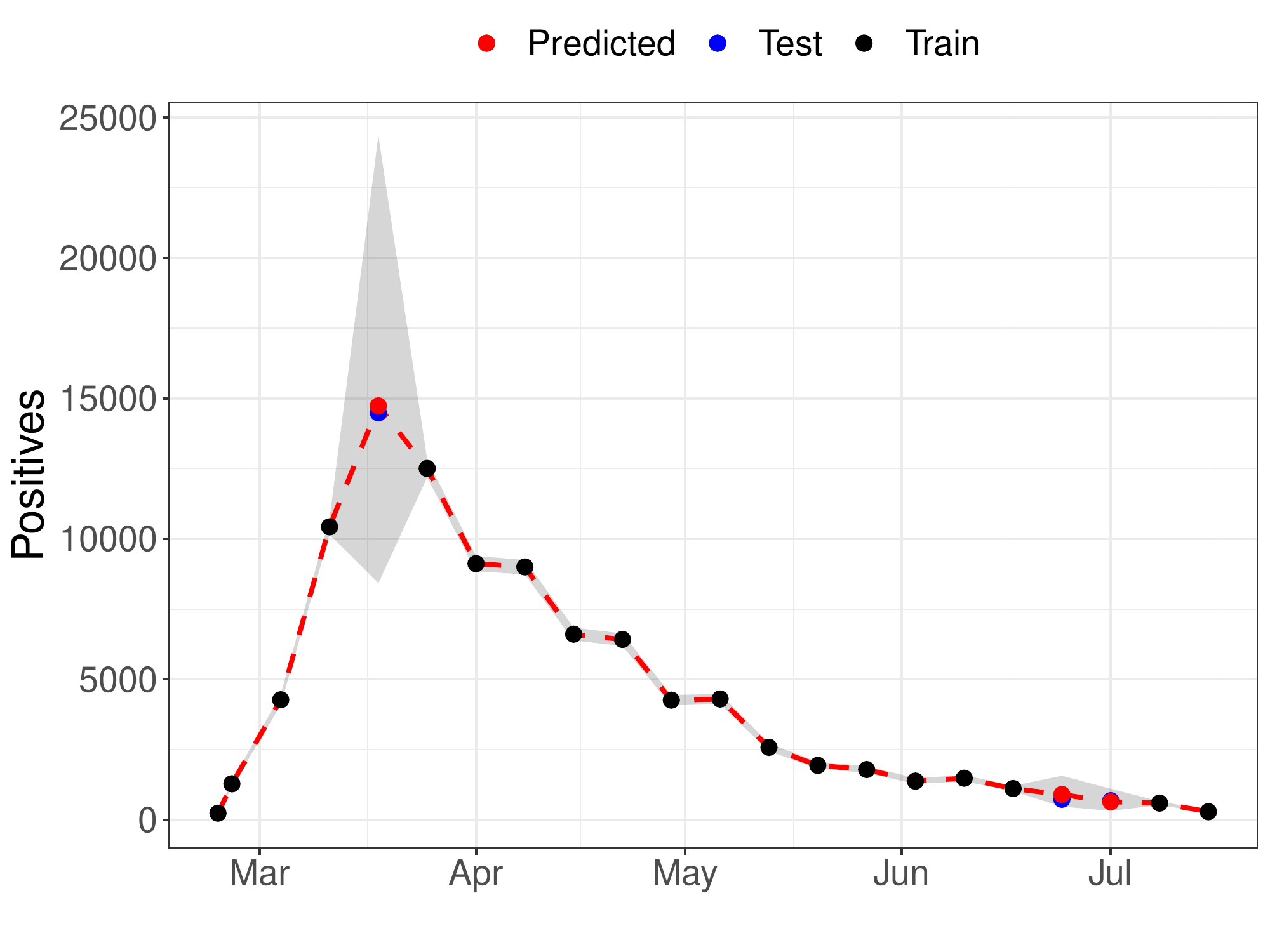}
    \caption{Lombardia}
    \label{fig:sardiniaI}
    \end{subfigure}
    \begin{subfigure}[b]{0.22\linewidth}
        \includegraphics[width = \textwidth]{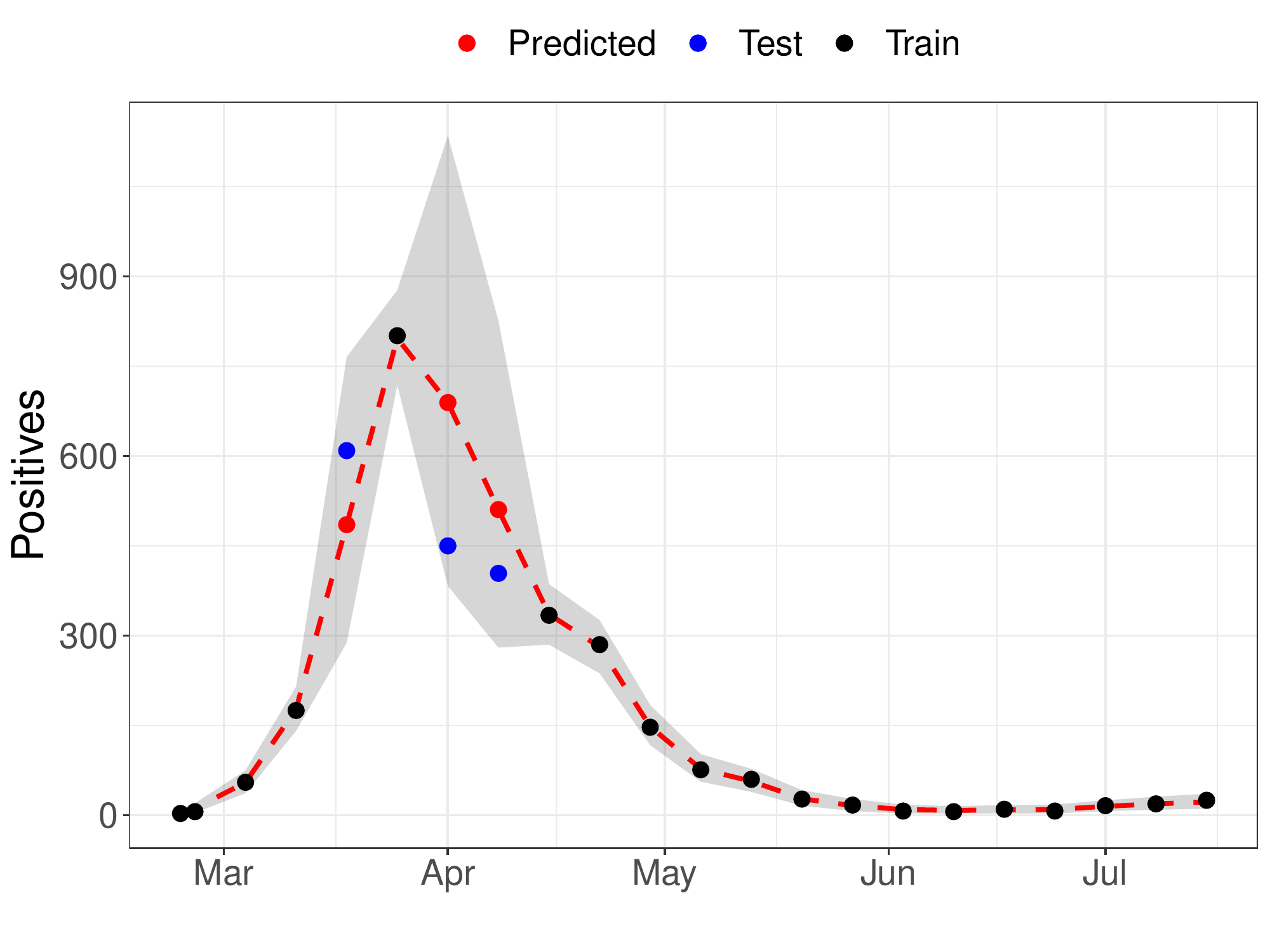}
    \caption{Sicilia}
    \label{fig:venetoI}
    \end{subfigure}
    \begin{subfigure}[b]{0.22\linewidth}
        \includegraphics[width = \textwidth]{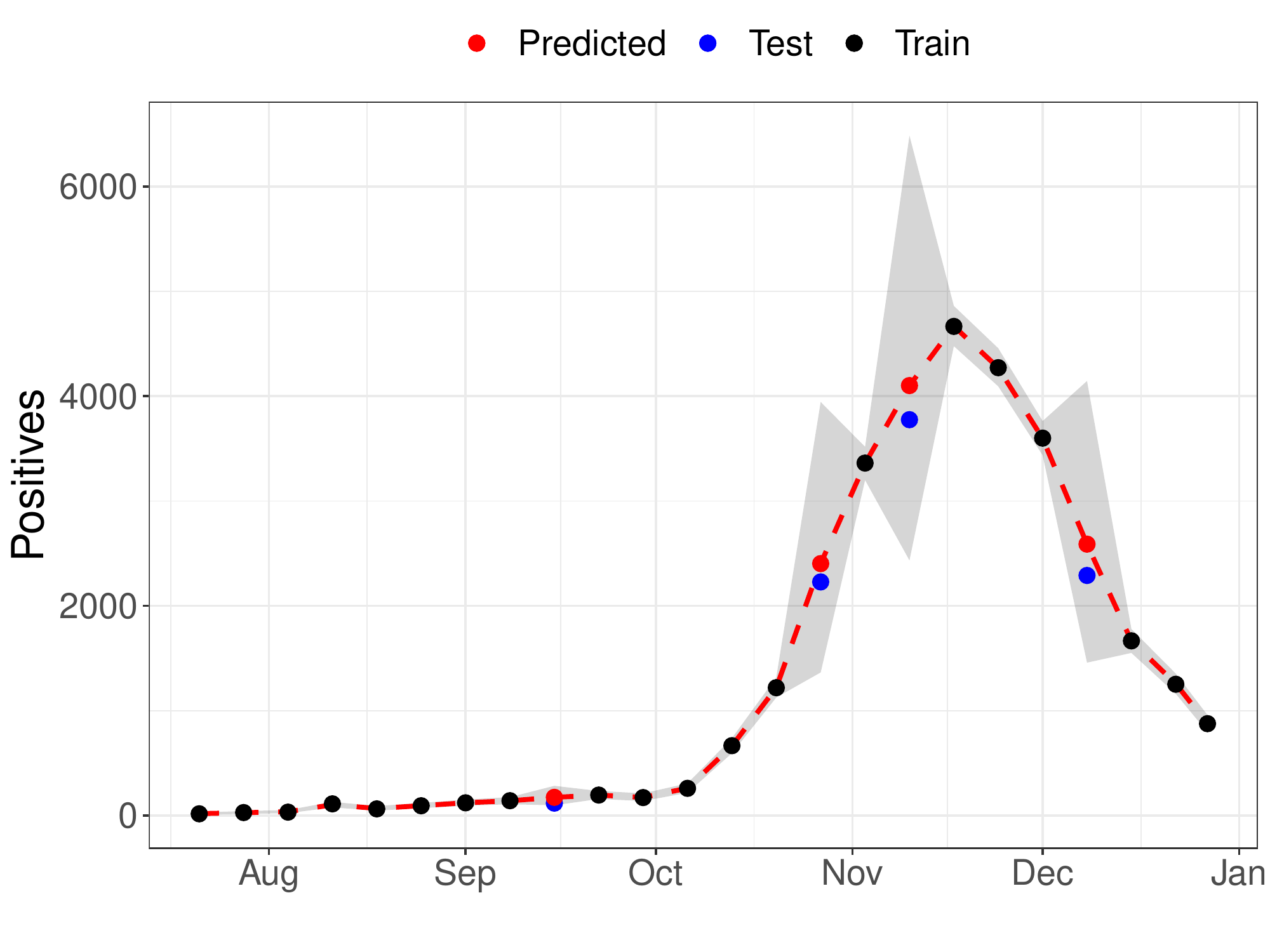}
    \caption{Abruzzo}
    \label{fig:moliseII}
    \end{subfigure}
    \begin{subfigure}[b]{0.22\linewidth}
        \includegraphics[width = \textwidth]{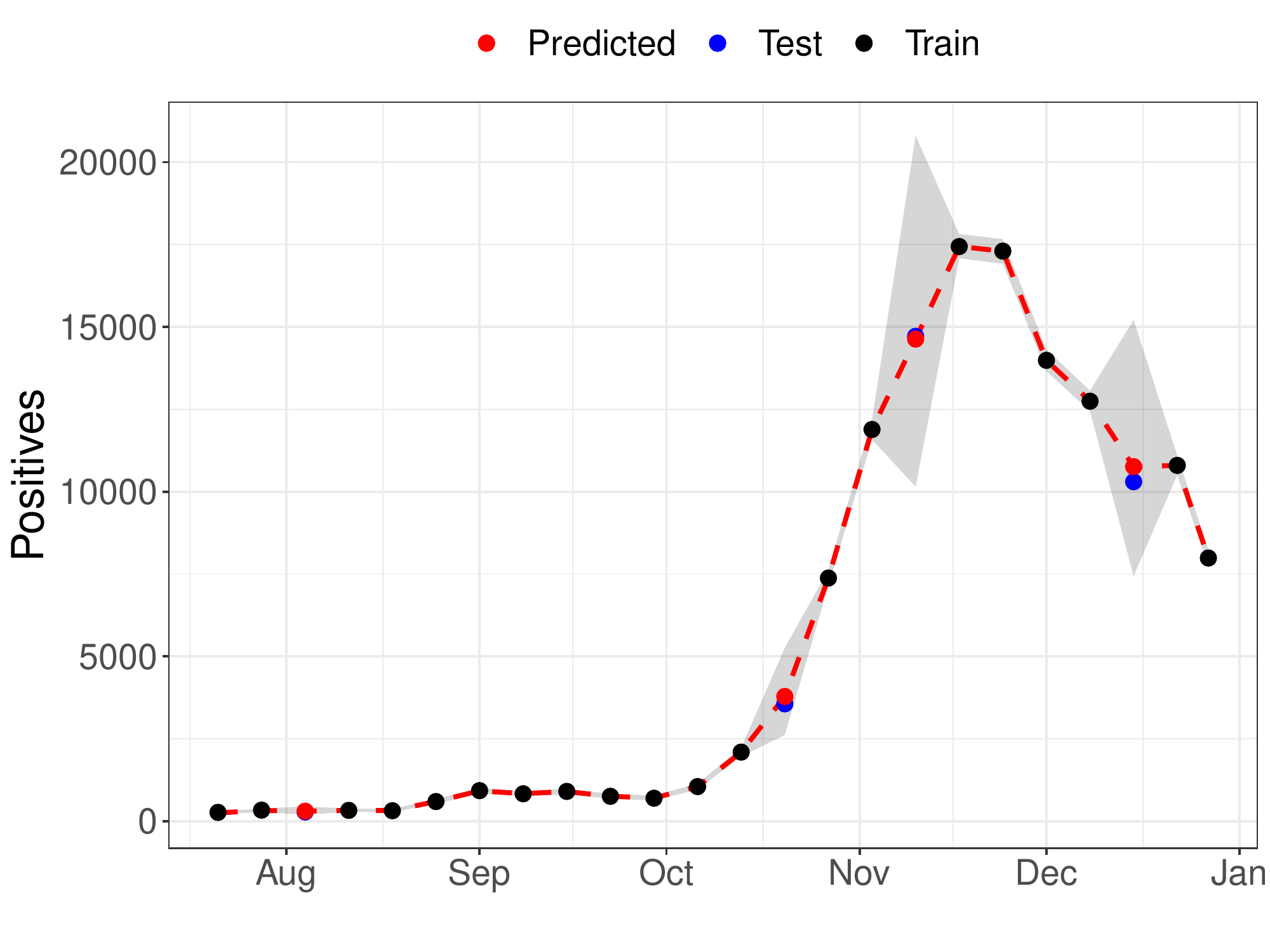}
    \caption{Emilia Romagna}
    \label{fig:lazioII}
    \end{subfigure}
    \begin{subfigure}[b]{0.22\linewidth}
        \includegraphics[width = \textwidth]{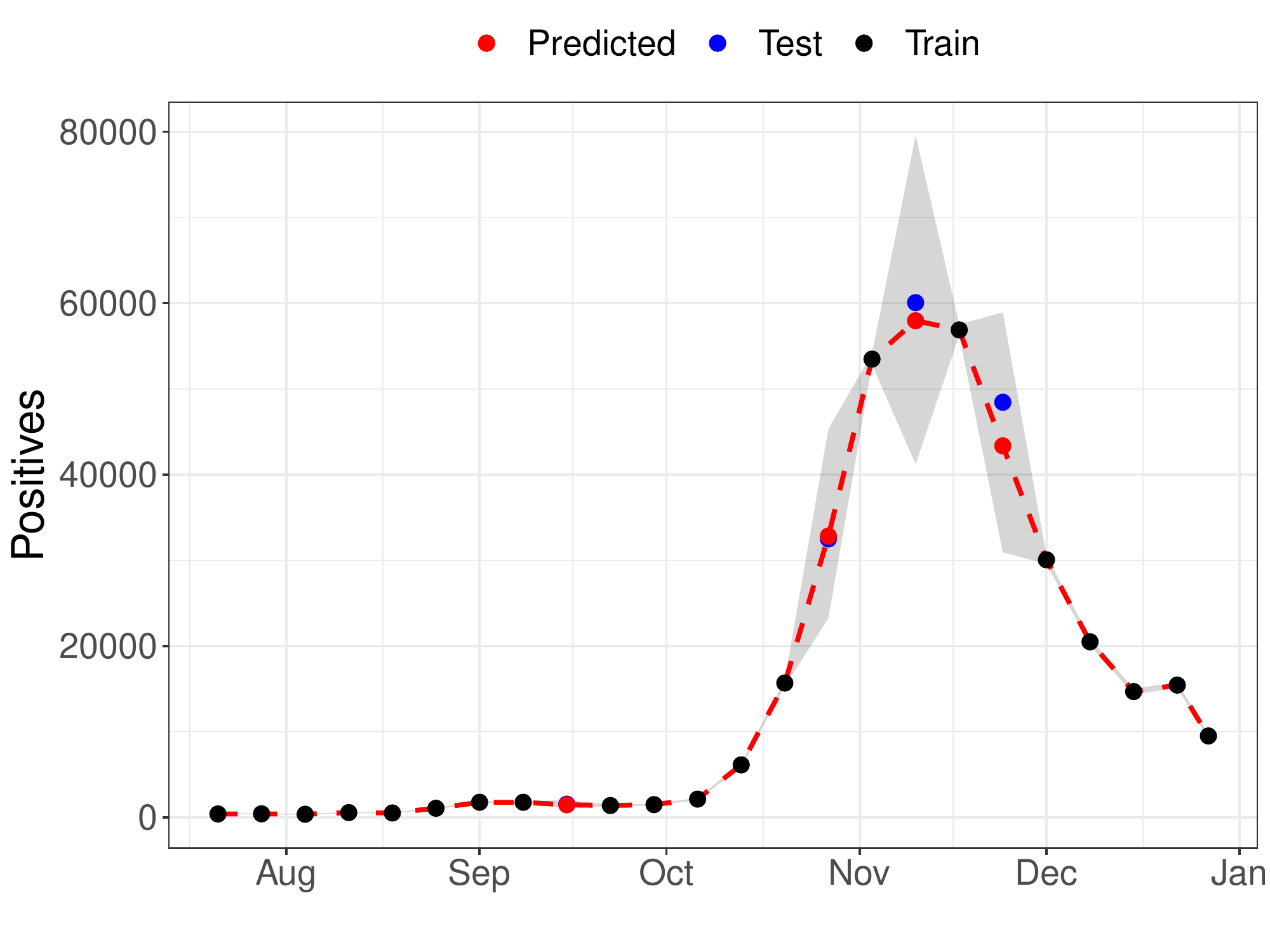}
    \caption{Lombardia}
    \label{fig:sardiniaII}
    \end{subfigure}
    \begin{subfigure}[b]{0.22\linewidth}
        \includegraphics[width = \textwidth]{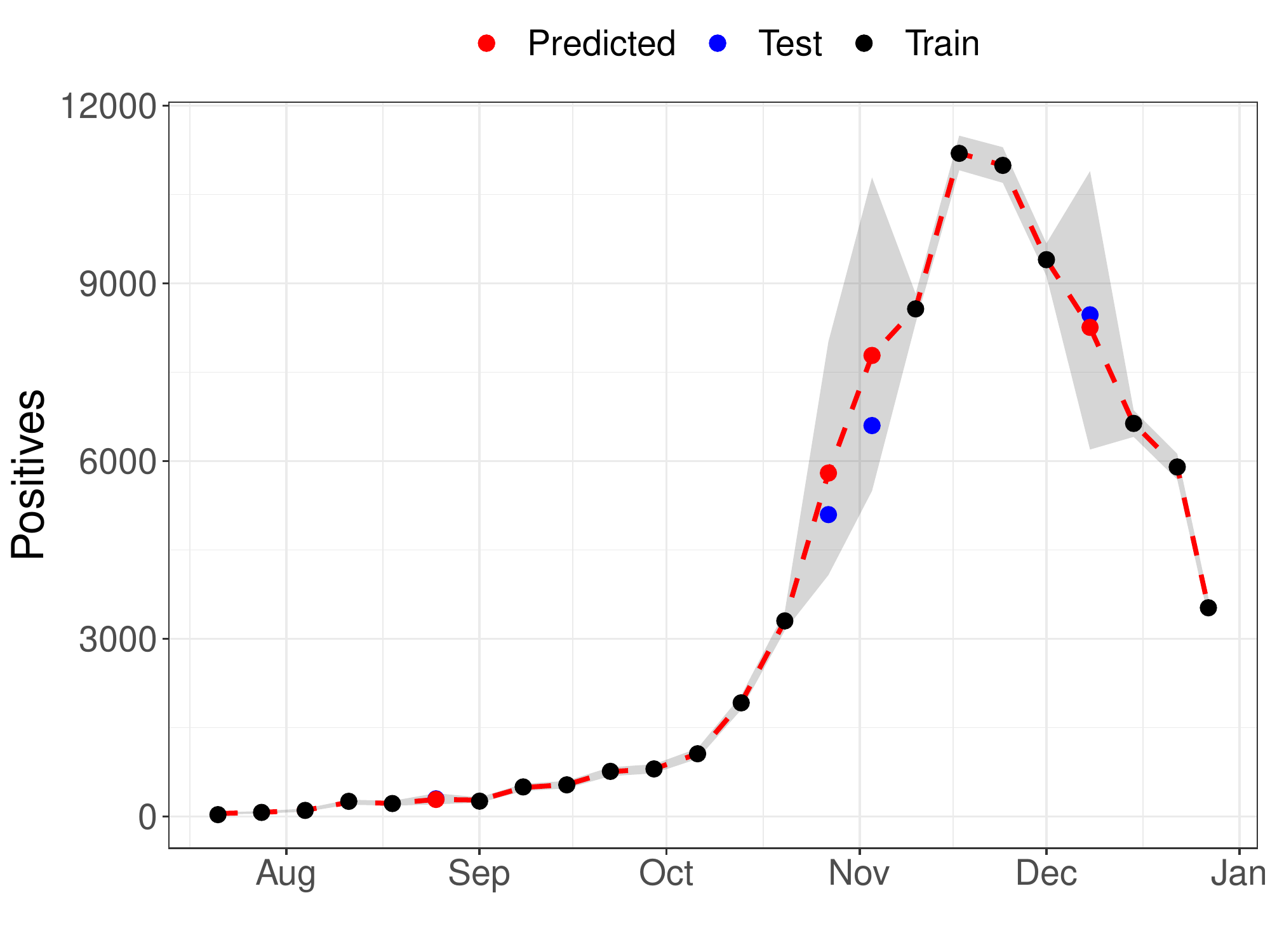}
    \caption{Sicilia}
    \label{fig:venetoII}
    \end{subfigure}
    \caption{Observed time series (red dots) and model fit (black solid) with $95\%$ prediction intervals (black dashed) for 4 randomly chosen regions in the first (top panels) and the second wave (bottom panels).}
    \label{modelfit}
\end{figure}

We argue that the chosen model is able to reconstruct the \textit{true evolution} of the incidence curve, also at missing points in the series. Some examples are in Fig.~\ref{fig:moliseI}-\ref{fig:venetoII}. The fit along all time points (in sample and out-of-sample) are plotted together with the $95\%$ posterior predictive intervals for four randomly selected regions (Abruzzo, Emilia Romagna, Lombardia and Sicilia), for the first and the second wave of the epidemic. 
We can clearly notice how the random effect allows the model to catch the wiggly behaviour of the observed data, and how almost all the observed data fall into the prediction intervals in spite of the large over-dispersion of the observed counts. More importantly, the predictive intervals obviously widen in correspondence of the missing observations, and practically always include the true value, even when this  deviates from a typical, expected behaviour (see again Figures~\ref{fig:moliseI}-\ref{fig:venetoII}). 
Given the homogeneity assumptions for Richards' curve for all regions, this must be mainly due to the dependence structure induced by $M_1$. 
Fig. \ref{fig:rmse} in the appendix shows specifically the out-of-sample predictive performances, where values in the test set are plotted on the log-scale.

\section{Discussion}
\label{sec:disc}

Modeling incident cases poses several issues, ranging from the discrete nature of the observations to the dependence structure of the data across times and proximity regions. The proposed generalised logistic growth curve accommodates the main data features, and provides a satisfactory solution for data analysis and prediction under a spatially heterogeneous framework. The proposal is applied to Italian regional data, but it can be applied to data from any other country. Similarly, if data were available at the province and/or municipality level, within-region spatial dependence could be explored, promptly identifying clusters of positive cases.

The spatio-temporal dependence is modeled through the inclusion of region-specific random effects.  The inclusion of random effects relaxes the working assumption used in \cite{di2020nowcasting}, where independence is assumed. Failure of the independence-assumed model to fit the data could be due to
misspecification of any of the elements defining the linear predictor. Here, not all possible covariates, such as population density or pollution exposure levels, were considered in model specification. Their joint effect is, however, summarized by latent variables, i.e. the random effects. On the one side, the additional computational burden can be dealt with in \texttt{Stan} within a Bayesian framework with minor efforts. On the other side, the improvements in the goodness of fit and predictions are evident.
As shown in Sec.~\ref{sec:results}, the dependence network plays a crucial role and gives useful insights. Here we analysed two separate waves in Italy. During the first wave, strict restrictions were applied, strongly limiting mobility across the country, mainly allowing for transportation routes only. As a result, the spread of the contagion was mainly influenced by geographic proximity, with northern regions being more affected by the epidemic than those in the Center and South of Italy. During summer, instead, people took the chance of less restrictive travelling constraints and enjoyed the summer season having holidays far from their  region of residence. This leads to a completely different spatial association. The country was not anymore divided into three geographical sections, but a more uniform development of the epidemic was observed. Regions coloring Fig.~5(b) reflects both the type of regional policy in terms of screening and the type of non pharmaceutical measures taken. For example Calabria region did not develop a consistent screening activity, and hence was subject to more strict restrictions than other regions.  

The latent dependence structure consistently aids interpretation. However, it may induce some bias in the predictions if the underling network is misspecified. To avoid such bias, the network might be explicitly modeled, and estimated together with all other parameters in the MCMC machinery. There are some examples of this approach in the recent literature \citep{rushworth2017adaptive, ejigu2020introducing, corpas2020use}, and it is a possible further development for our model.
In the future we will also consider weighted spatial structures as in \cite{della2020network}, by specifying $\bW_1$ as $\lrnd\bW_1 +\bW_1^{\top}\rrnd/2$. This was not pursued in this paper in order 
to directly compare unweighted versions of the two spatial graphs. 

An important extension would be the development of a space-time model capable to capture the entire evolution of an epidemic; fitting all waves within the same model specification. This might be done by developing a model based on a mixture of Richards' curves, each one capable to describe individual epidemic waves. 

Further results (e.g. chains diagnostics) are available from the authors upon request and we point again the reader to the public \texttt{GitHub} repository available at \url{https://github.com/minmar94/Covid19-Spatial}.


	\bibliographystyle{plainnat}
    \bibliography{template}
\newpage
\section*{Appendix}
\begin{figure}[H]
    \centering
    \begin{subfigure}[b]{0.45\linewidth}
    \centering
    \includegraphics[width = 0.45\textwidth]{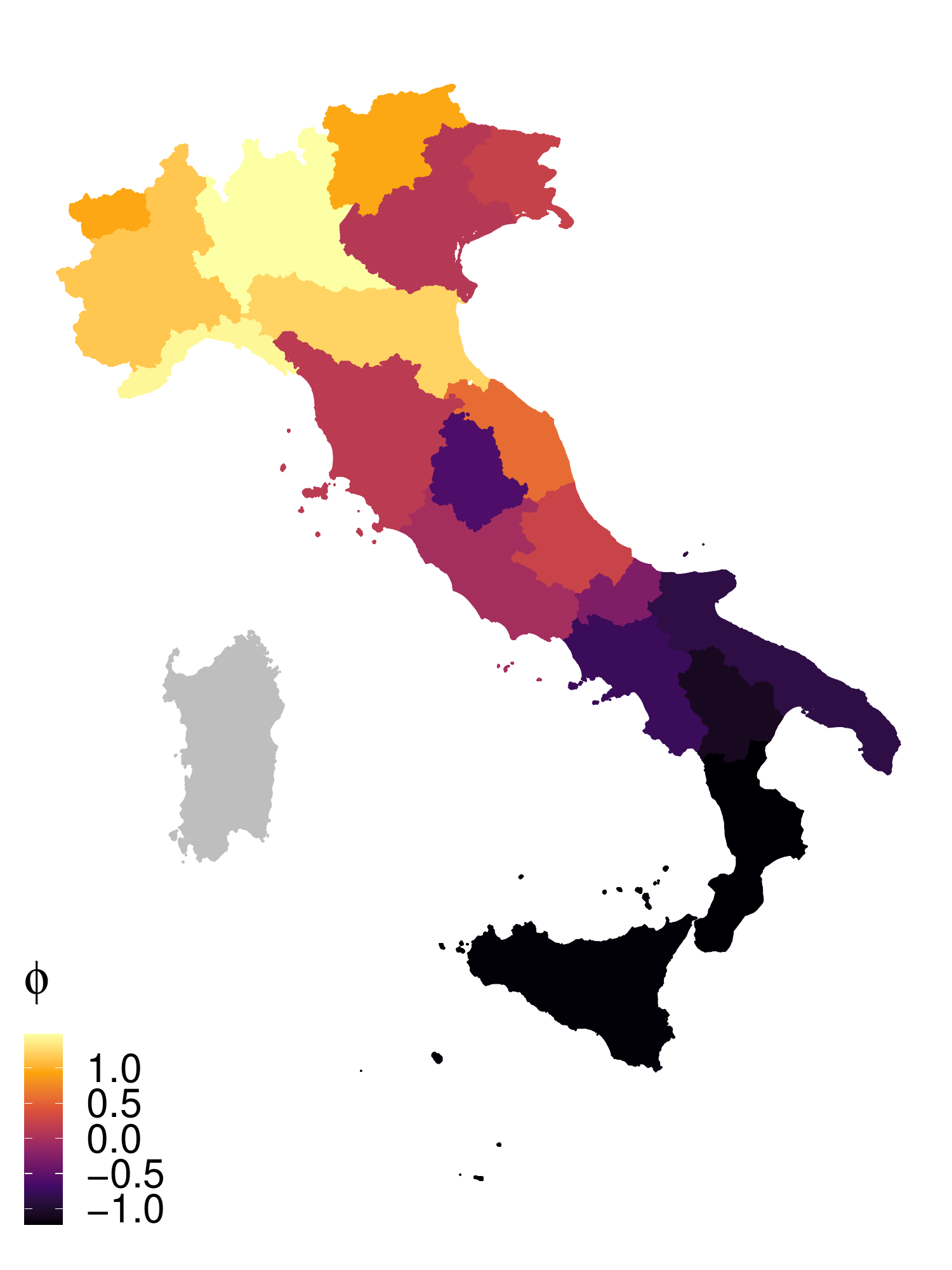}
    \caption{}
    \label{fig:PhiGeo}
    \end{subfigure}
    \hspace{0cm}
    \begin{subfigure}[b]{0.45\linewidth}
    \centering
    \includegraphics[width = 0.45\textwidth]{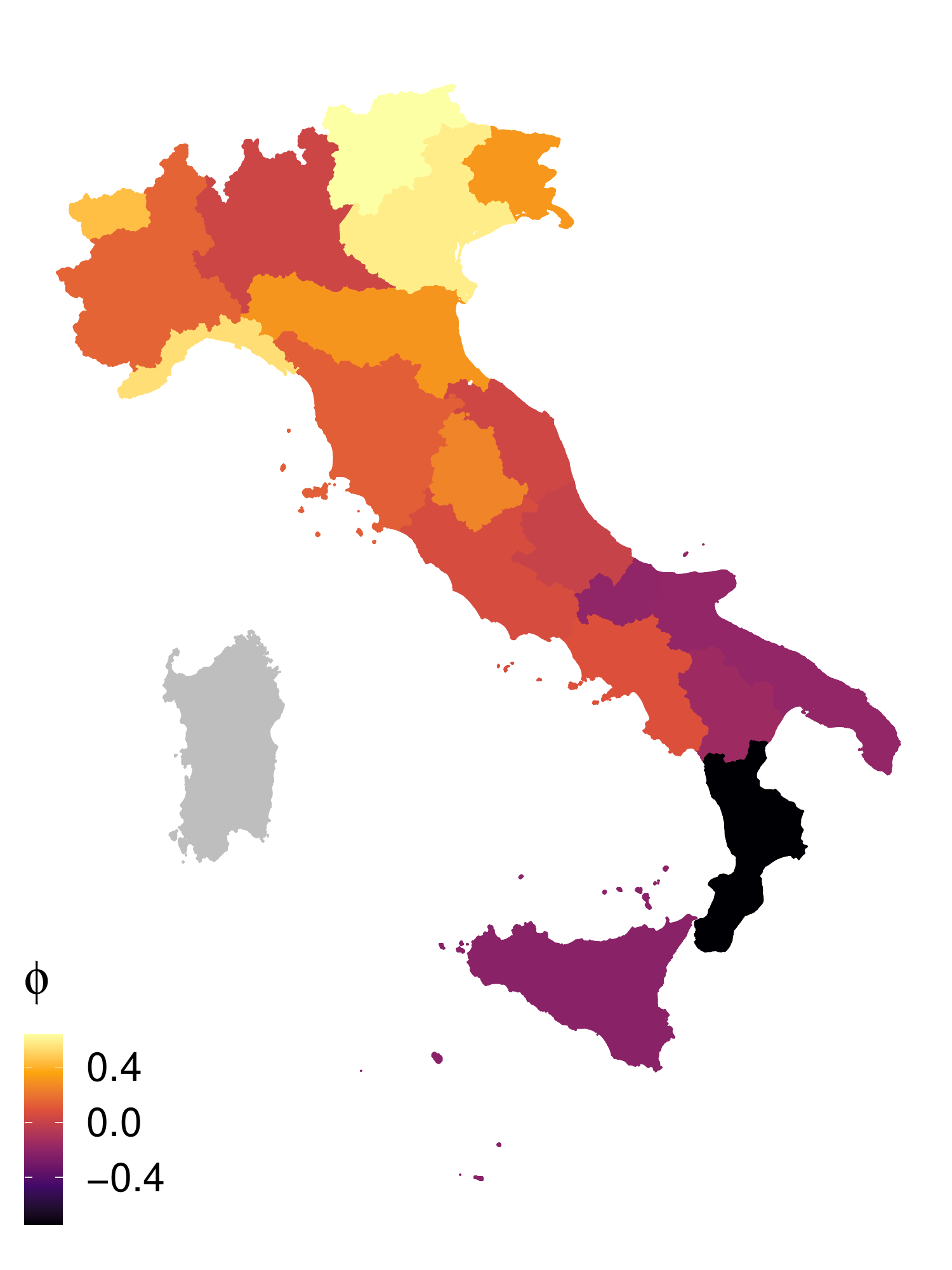}
    \caption{}
    \label{fig:PhiGeoII}
    \end{subfigure}
    \caption{Average estimated spatial random effect $\bar{\phi}_g$ by $M_2$ for the first (a) and the second (b) wave.}
\end{figure}

\begin{figure}[H]
    \centering
    \begin{subfigure}[b]{0.9\linewidth}
        \includegraphics[width = \textwidth]{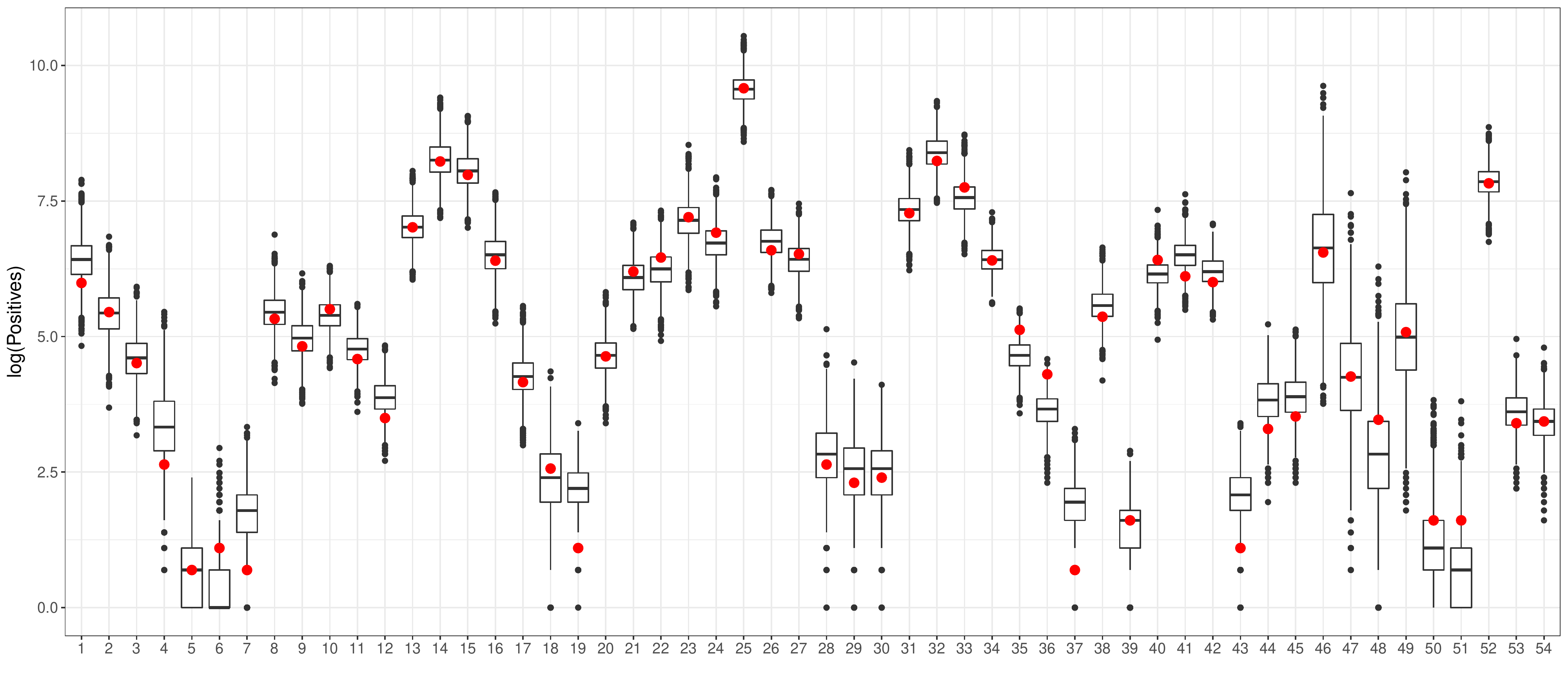}
    \caption{}
    \label{fig:rmseI}
    \end{subfigure}
     \begin{subfigure}[b]{0.9\linewidth}
        \includegraphics[width = \textwidth]{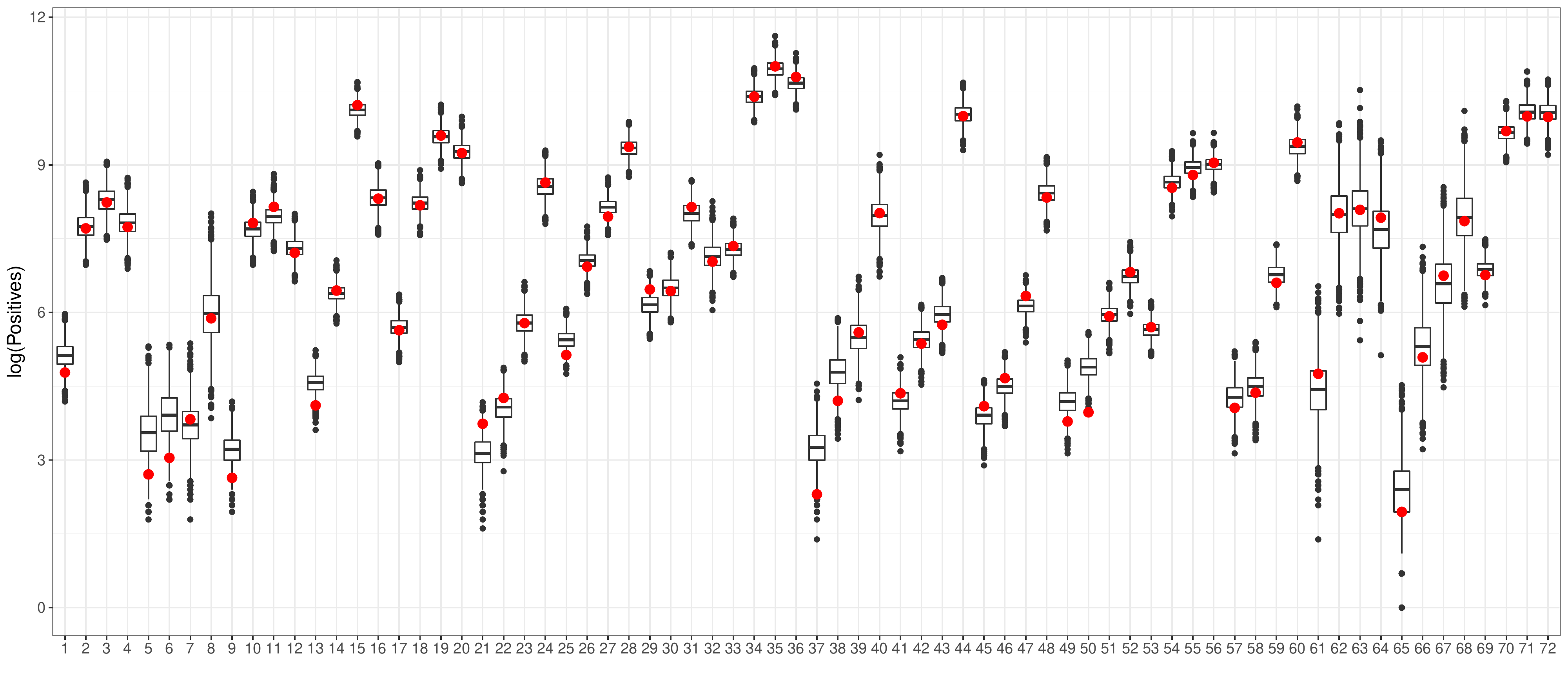}
    \caption{}
    \label{fig:rmseII}
    \end{subfigure}
    \caption{Observed points (red) and simulated predictions (boxplots) for the first (a) and the second wave (b) test sets.}
    \label{fig:rmse}
\end{figure}

\end{document}